\UseRawInputEncoding

\documentclass[11pt]{article}
\usepackage{}
\usepackage{amssymb}
\usepackage{amsfonts}
\usepackage{mathrsfs}
\usepackage{graphicx}
\usepackage{amsmath}
\usepackage{color}
\usepackage{amsthm}
\usepackage{epstopdf}
\usepackage{subfigure}
\usepackage{pifont,bm}
\usepackage{tikz}
\usepackage{enumerate}
\usepackage{mathrsfs}
\usepackage{cite}
\usepackage{hyperref}
\usepackage{booktabs}
\usepackage{diagbox}
\usepackage{makecell}

\usepackage{appendix}

\theoremstyle{definition}

\makeatletter
\def\@biblabel#1{[#1]}
\makeatother

\makeatletter \@addtoreset{equation}{section}

\begin{document}

\begin{titlepage}
\title{\bf{Complex dynamics on the one-dimensional quantum droplets via time piecewise PINNs
\footnote{Corresponding authors.\protect\\
\hspace*{3ex} E-mail addresses: ychen@sei.ecnu.edu.cn (Y. Chen)}
}}
\author{Juncai Pu$^{a}$, Yong Chen$^{a,b,*}$\\
\small \emph{$^{a}$School of Mathematical Sciences, Shanghai Key Laboratory of Pure Mathematics and} \\
\small \emph{Mathematical Practice, East China Normal University, Shanghai, 200241, China} \\
\small \emph{$^{b}$College of Mathematics and Systems Science, Shandong University }\\
\small \emph{of Science and Technology, Qingdao, 266590, China} \\
\date{}}
\thispagestyle{empty}
\end{titlepage}
\maketitle

\vspace{-0.5cm}
\begin{center}
\rule{15cm}{1pt}\vspace{0.3cm}

\parbox{15cm}{\small
{\bf Abstract}\\
\hspace{0.5cm}

The dynamics of one-dimensional quantum droplets and the landing applications of deep learning are recent research hotspots. In this work, we propose a novel time piecewise physics-informed neural networks (PINNs) to study complex dynamics on the one-dimensional quantum droplets by solving the corresponding amended Gross-Pitaevskii equation. The training effect of this network model in the long time domain is far better than that of the conventional PINNs, and each of its subnetworks is independent and highly adjustable. By using time piecewise PINNs with scarce training points, we not only study intrinsic modulation of single droplet and collision between two droplets, but also excite the breathers on droplet background. Intriguingly, we obtain an interference pattern from training result of collision between two droplets, which is a significant feature of the interplay of coherent matter waves. The numerical results showcase that different parameters may lead to completely different dynamic behaviors under the same initial condition in a nonlinear non-integrable system. Our results provide the significant guidance for intrinsic modulation of single droplet, droplet collision and breathers excitation via deep learning technology.

}

\vspace{0.5cm}
\parbox{15cm}{\small{

\vspace{0.3cm} \emph{Key words: quantum droplets; time piecewise physics-informed neural networks; Gross-Pitaevskii equation; breathers}  \\

} }
\end{center}
\vspace{0.3cm} \rule{15cm}{1pt} \vspace{0.2cm}

\section{Introduction}
Deep learning has played an increasingly substantial role in a wide range of research fields, and has achieved remarkable success in various applications \cite{LeCun-N2015}. However, its successful application in solving partial differential equation (PDE) has emerged only recently \cite{Karniadakis-NRP2021}. Recently, a network architecture based on deep learning was proposed, namely physics-informed neural network (PINN) \cite{Raissi-JCP2019}, it solve the PDE by embedding the PDE into neural network and using back-propagation based on the automatic differentiation to optimic loss function \cite{Baydin-JMLR2018,Rumelhart-N1986}. As an alternative approach to the traditional numerical PDE solver, the PINN algorithm is mesh-free and simple in flow, it can not only be applied to different types of PDEs in a wide range of forward problems when the amount of data is limited or when the problem is ill-posed, including fractional PDEs \cite{Pang-SIAMJSC2019}, integro-differential equations \cite{Lu-SIAMR2021}, radiative transfer equation \cite{Mishra-JQS2021} and stochastic PDEs \cite{Zhang-SIAMJSC2020}, but also easy to solve inverse problems of PDEs. PINN has been successfully applied to solve various problems in different fields, such as optics \cite{Chen-OE2020}, thermodynamics \cite{Cai-JHT2020}, fluid mechanics \cite{Raissi-S2020}, material mechanics \cite{Zhang-SA2022}, systems biology \cite{Yazdani-PCB2020}, aerodynamics \cite{Mao-CMAME2020} and biomedicine \cite{Costabal-FP2020}. Hence, the successful application of PINN in aforementioned fields reminds us whether PINN algorithm can be successfully applied to the study of complex dynamics of quantum droplet (QD)?

Ultracold Bose gas, which benefits from its macroscopic quantum properties, high controllability and experimental observability, is one of the ideal platforms for studying nonlinear waves and simulating various quantum states \cite{Anderson-S1995,Davis-PRL1995}. The realization of QD is one of the important recent achievements in the study of ultracold Bose gases and superfluids, and has made considerable progress in theoretical and experimental research \cite{Petrov-PRL2015,Petrov-PRL2016,Barbut-PRL2016,Cabrera-S2018}.
It is well known that a multi-dimensional Bose-Einstein condensates (BECs) with attraction between atoms experiences collapse \cite{Pethick-CUP2011}. Quantum fluctuations can arrest the collapse in two and three dimensions, such a possibility benefits from the expression of a correction, known as the Lee-Huang-Yang (LHY) term \cite{Lee-PR1957}, to the ground state energy of Bose gas due to quantum fluctuations, this correction corresponds to the appearance of the effective repulsion. QDs are small clusters of atoms self-bound by the balance of the balance of the gravitation between atoms and the effective repulsive force caused by quantum fluctuations. The existence of QDs in theory predicted in Ref. \cite{Petrov-PRL2015}, and QDs were also observed in strongly dipolar Bose gas \cite{Barbut-PRL2016} and mixtures of BECs \cite{Cabrera-S2018} experimentally. Thus we consider using the PINN stem from deep learning to numerically study the complex dynamic behaviors of one-dimensional QDs with a small amount of data on computer.

The celebrated Gross-Pitaevskii equation (GPE) describing one-dimensional QDs provides a complete analysis of their static characteristics, such as shape and energy \cite{Pitaevskii-OUP2016}. GPE was derived independently by Gross \cite{Gross-NC1961} and Pitaevskii \cite{Pitaevskii-ZETF1961} and is the main theoretical tool for investigating nonuniform dilute Bose gases at low temperature. Furthermore, the dynamic behavior of matter wave soliton and the excitation mechanism of matter wave breather are consistent with the theoretical prediction of GPE \cite{Astrakharchik-PRA2018,Lv-PLA2022}. Chen research team has studied data-driven nonlinear waves for several nonlinear systems using PINN, such as nonlinear Schr\"{o}dinger equation (NLS) \cite{Pu-CPB2021}, Chen-Lee-Liu equation \cite{Peng-CNSNS2021} and Manakov systems \cite{Pu-CSF2022}, and they have adopted a variety of strategies to improve PINN for improving training effect and network performance, including conservation law \cite{Lin-JCP2022}, adaptive activation function \cite{Pu-ND2021}, nonlocal operator \cite{Peng-PD2022}, parameter regularization \cite{Pu-YO2021}, etc. However, previous PINN architectures have excellent effect on the data driven forward-inverse problems of nonlinear systems in a short time interval, while the matter wave dynamics and excitation behavior on one-dimensional QDs need to be studied in a long time domain. Therefore, this paper we propose a novel PINN model, that is time piecewise PINNs (tpPINNs), we segment the temporal domain, then use multiple consecutive subnet of PINNs, where each subnet has a freely adjustable number of subnetwork layers, initial/boundary data points, remaining collocation points, adaptive activation function weights, and parameter regularization weights, so as to obtain better global training effect.

In this work, we utilize tpPINNs with scarce data to address intrinsic oscillations of an isolated droplet, collisions between two QDs and the breather excitation on one-dimensional QDs. The paper is organized as follows. In section 2, we present the novel tpPINNs model and showcase the well posed conditions for the existence of QDs in the GPE. Section 3 displays numerical results, including the intrinsic modulation of a single droplet, the collision dynamics of two droplets and the breather excitation on one-dimensional droplet. We summarize the conclusions of our work in section 4.

\section{Methodology and model system}
In real-world applications, we usually encounter the following situations: the model studied is uncertain, data capture is very costly or the data itself is sparse in nature. Generally, we consider a multi-dimensional spatiotemporal real nonlinear system whose governing equations can be described by a set of coupled, parameterized and variable coefficient PDEs in the general form given by
\begin{align}\label{E1}
\bm{q}_t+\mathcal{F}[\bm{q},\bm{q}^2,\cdots,\nabla_{\bm{\mathrm{x}}}\bm{q},\nabla^2_{\bm{\mathrm{x}}}\bm{q},\cdots,\bm{q}\cdot\nabla_{\bm{\mathrm{x}}}\bm{q},\cdots;\bm{\lambda}]=0,
\end{align}
in which $\bm{q}=\bm{q}(\bm{\mathrm{x}},t)\in\mathbb{R}^{1\times n}$ is the $n$-dimensional latent solution, $t\in[0,T]$ denotes time and $\bm{\mathrm{x}}\in\Omega$ specifies the space, while $\bm{q}_t$ is the first-order time derivative term, $\nabla$ is the gradient operator with respect to $\bm{\mathrm{x}}$, $\mathcal{F}[\cdot]$ is a complex nonlinear operator of $\bm{q}$ and its spatial derivatives, $\bm{\lambda}$ parameterizes nonlinear system via arbitrary constants or provides variable coefficient by selecting arbitrary functions of $t$ and $\bm{\mathrm{x}}$. Then, we also consider the initial and boundary conditions of spatiotemporal nonlinear system denoted by
\begin{align}
&\qquad\mathcal{I}[\bm{q};\bm{\mathrm{x}}\in\Omega,t=0]=0,\nonumber\\
&\mathcal{B}[\bm{q},\nabla_{\bm{\mathrm{x}}}\bm{q};\bm{\mathrm{x}}\in\partial\Omega,t\in[0,T]]=0.\nonumber
\end{align}
If we consider a $\bm{\hat{q}}$ of nonlinear system in the complex analytic space, we can derive two real-value functions $\bm{\hat{u}}$ and $\bm{\hat{v}}$ in two real nonlinear system spaces by using transformation $\bm{\hat{q}}=\bm{\hat{u}}+\mathrm{i}\bm{\hat{v}}$. Different from the traditional numerical methods, for studying the initial-boundary value problems of such nonlinear systems \eqref{E1}, PINN has a powerful, simple and accurate training effect in the short time domain. Since we consider the QDs in BECs, which leads to the possibility that the upper limit $T$ of the time variable may be large, so that it is very difficult to train via the traditional PINN, thus we propose a novel PINN model for studying for a long time domain in this work.

\subsection{Time piecewise PINNs}
In order to study the data driven problem of nonlinear systems in a long time domain under the initial-boundary value condition, a natural inspiration is to decompose the time domain, conduct segmented training through multiple PINNs, and then reorganize the training results, so as to obtain excellent training effects in the entire time domain. For convenience, we call the entire network model is ``time piecewise PINNs", and define the PINN of each time domain is a subnet. Meanwhile, each subnet is independent and highly adjustable, namely it can be trained in any time domain, and can choose arbitrary super parameter, number of network layers and neurons, etc.

Indeed, the PINN model with neuron-wise locally adaptive activation functions and slope recovery term can improve the convergence speed, stability of the loss function and approximation ability in the training process from Ref. \cite{Jagtap-PRSA2020}. Moreover, $L^2$ parameter norm penalty can modify the weights and reduce the impact of large weights on the network, choosing a parameter regularization strategy with appropriate weights can improve the performance of PINN in studying the forward and inverse problems of nonlinear systems \cite{Pu-YO2021}. Therefore, we choose each subnet as a PINN with adaptive activation function and parameter regularization to enhance time localized training performance.

For simplicity, we reformulate Eq. \eqref{E1} in the following
\begin{align}\label{E2}
\bm{q}_t+\bm{Q\Lambda}=0,
\end{align}
where $\bm{Q=Q(q)}\in\mathbb{R}^{1\times m}$ is an extensive library of symbolic functions consisting of many candidate terms, such as constant, polynomial, and trigonometric terms with respect to each spatial dimension. $m$ denotes the total number of candidate terms in the library. $\bm{\Lambda}\in\mathbb{R}^{m\times n}$ is the sparse coefficient matrix, one can contains arbitrary constants or arbitrary functions of $t$ and $\bm{\mathrm{x}}$. Thus, the data-driven problem can then be stated as: given the spatiotemporal measurement initial and boundary data, obtain solutions and find sparse $\bm{\Lambda}$ such that Eq. \eqref{E2} holds. Then for conveniently introducing the innovative tpPINNs paradigm, we assume $\bm{q}=\{u,v\}$ and $\bm{\mathrm{x}}=\{x\}$.

\begin{figure}[htbp]
\centering
\begin{minipage}[t]{0.99\textwidth}
\centering
\includegraphics[height=6cm,width=12cm]{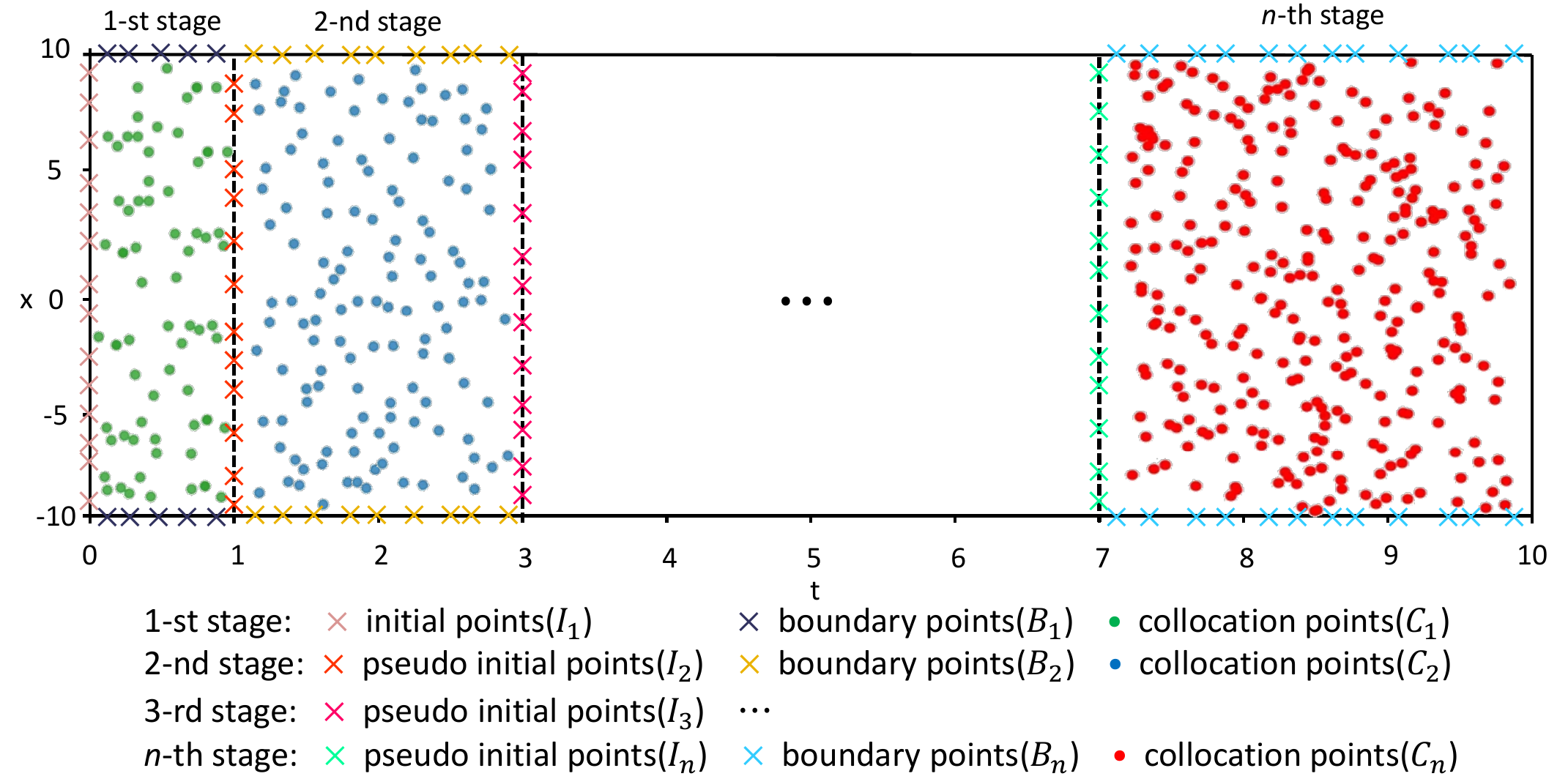}
\end{minipage}
\centering
\caption{(Color online) the example distribution of sampling points at each stage in the tpPINNs. For convenience, where we select $t_1=1,t_2=3,t_{n-1}=7, T=10,x_{\rm lb}=-10,x_{\rm ub}=10.$}
\label{F1}
\end{figure}

In tpPINNs algorithm, the boundary points [the locations of sampling points for the upper/lower boundary are consistent in the same time domain] are divided into $n$ parts according to the length of time domain. This means the algorithm has $n$ stages PINNs for training, and each stage uses a subset of boundary points. Using the boundary points to generate $n$ small subsets $\Big\{B_k=\big\{(x_{\rm{b}},t^j_{\mathrm{b},k})\big\}^{N^k_{\rm{b}}}_{j=1}\Big|k=1,\cdots,n,[0,t_{1}]\bigcup[t_{1},t_{2}]\bigcup\cdots\bigcup[t_{n-1},T]=[0,T](t_0=0,t_n=T),t^j_{\mathrm{b},k}\in[t_{k-1},t_{k})\Big\}$, where $N^k_{\mathrm{b}}$ denotes the number of training points on the boundary condition of the $k$-th PINN, $x_{\mathrm{b}}$ contains upper and lower boundary $x_{\mathrm{ub}}$ and $x_{\mathrm{lb}}$. We define the $n$ stages PINNs all initial points $\Big\{I_k=\big\{(x^l_{\rm{i}},t_{\mathrm{i},k})\big\}^{N^k_{\rm{i}}}_{l=1}\Big|k=1,\cdots,n,t_{\mathrm{i},1}=0\Big\}$, in which $N^k_{\rm{i}}$ represents the number of training points on the initial condition of the $k$-th PINN, as well as we define $\{I_1\}$ is initial points and $\{I_k (k\geqslant2)\}$ are pseudo initial points, the pseudo initial points of each stage generated from the training results of the previous stage PINN. That is, we only provide the initial points $\{I_1\}$ of the 1-st PINN for training, then the pseudo initial points of the 2-nd stage PINN are automatically derived from the training result of the 1-st stage PINN, and the pseudo initial points of the $k$-th ($k\geqslant2$) stage PINN are automatically derived from the training result of the $(k-1)$-th stage PINN. The collocation points $\Big\{C_k=\big\{(x^r_{\mathrm{c},k},t^r_{\mathrm{c},k})\big\}^{N^k_{\rm{c}}}_{r=1}\Big|k=1,\cdots,n\Big\}$ of each stage are arbitrarily selected, where $N^k_{\rm{c}}$ expresses the number of collocation points in the $k$-th stage PINN. Intuitively, the example distribution of sampling points in each stage is displayed in the Fig. \ref{F1}. In which, the initial points (or pseudo initial points) and boundary points of each stage are shown in detail by using different color marks ``$\times$'', then the collocation points of different stages are distinguished by circle dots with different colors.

From Fig. \ref{F1}, one can know that tpPINNs model, which contains $n$ stage PINNs, is trained sequentially through training points [including initial points, pseudo initial points, boundary points and collocation points] in chronological order, and each stage PINN with adaptive activation function and parameter regularization is similar. Thus we only detailed introduce the $k$-th stage PINN in this part. We establish a $D$ layers neural network (NN) with an input layer, $D-1$ [also called $L$] hidden-layers and an output layer, in which the $d$-th hidden-layer contain $N_d$ number of neurons. Each NN hidden-layer accepts an output $\bm{\mathrm{x}}^{d-1}\in\mathbb{R}^{N_{d-1}}$ from the previous layer, where an affine transformation can be yielded as follows form
\begin{align}\label{E3}
\mathcal{A}_d(\bm{\mathrm{x}}^{d-1})\triangleq\bm{W}^d\bm{\mathrm{x}}^{d-1}+\bm{b}^d,
\end{align}
in which the weights and bias term are $\bm{W}^{d}\in\mathbb{R}^{N_d\times N_{d-1}}$ and $\bm{b}^d\in\mathbb{R}^{N_d}$ in the NN associated with the $d$-th layer. Then we introduce the neuron-wise locally adaptive activation functions into $d$-th layer NN, it defined as bellow
\begin{align}\nonumber
\sigma\Big(Sa^d_i\big(\mathcal{A}_d(\bm{\mathrm{x}}^{d-1})\big)_i\Big),\,i=1,2,\cdots,N_d,
\end{align}
where $\sigma$ is the activation function, $S\geqslant1$ are scaling factors and $\{a^d_i\}$ are additional $\sum\limits_{d=1}^{D-1}N_d$ parameters to be optimized. Whereas, There exists a critical scaling factor $S_{c}$ above which the NN optimization algorithm becomes very sensitive for every problem \cite{Jagtap-PRSA2020}. Neuron-wise locally activation function acts as a vector activation function in each hidden-layer, where every neuron has its own slope for the activation function. The resulting optimization problem leads to finding the minimum of a loss function by optimizing activation slopes along with weights and biases. Then the final NN representation of the solution based on the neural-wise locally adaptive activation function is manifested as following
\begin{align}\label{E4}
\bm{q}(t,\bm{\mathrm{x}};\bm{\theta})=\Big(\left(\mathcal{A}_D\right)_{i'}\circ\sigma\circ Sa^{D-1}_{i}\left(\mathcal{A}_{D-1}\right)_{i}\circ\cdots\circ\sigma\circ Sa^1_i\left(\mathcal{A}_1\right)_i\Big)(\bm{\mathrm{x}}),\,i'=n.
\end{align}
where $t$ and $\bm{\mathrm{x}}$ denote the the inputs, and $\bm{q}(t,\bm{\mathrm{x}};\bm{\theta})$ represent outputs of the NN. $n$ is dimension of $\bm{q}$. The set of trainable parameters $\bm{\theta}\in\mathcal{N}$ consists of $\big\{\bm{W}^d,\bm{b}^d\big\}_{d=1}^{D}$ and $\big\{a_i^d\big\}_{d=1}^{D-1},\forall i=1,2,\cdots,N_d$, in which $\mathcal{N}$ is the parameter space [if consider the inverse problem, $\mathcal{N}$ also includes parameters $\bm{\theta}$ to be predicted].

For constructing physics-informed part of PINN, we introduce the physical constraint of solution $\bm{q}$ via the nonlinear system \eqref{E1}. Thus the deformation Eq. \eqref{E2} constitute the physics-informed parts of the NN, which can be defined as
\begin{align}\label{E5}
\begin{split}
&f_{\bm{q}}:=\bm{q}_t+\bm{Q\Lambda}.\\
\end{split}
\end{align}

The main characteristic of the PINN is that it can readily incorporate all the given information like governing equation, initial/boundary conditions, experimental data, slope recovery term, parameter regularization etc into the loss function thereby recast the original problem into an optimization problem. The PINN algorithm aims to learn a surrogate $\bm{q}^{\bm{\theta}}=\bm{q}(t,\bm{\mathrm{x}};\bm{\theta})$ for predicting the solution $\bm{q}$ of the governing PDE. In $k$-th stage PINN algorithm, the total loss function is defined as
\begin{align}\label{E6}
\mathcal{L}(\mathcal{D}_{\rm{i}},\mathcal{D}_{\rm{b}},\mathcal{D}_{\rm{c}};\bm{\theta,\Lambda})=\mathcal{L}_{\rm{id}}(\mathcal{D}_{\rm{i}};\bm{\theta})+\mathcal{L}_{\rm{bd}}(\mathcal{D}_{\rm{b}};\bm{\theta})+\mathcal{L}_{\rm{r}}(\mathcal{D}_{\rm{c}};\bm{\theta,\Lambda})+\frac{\alpha}{2}\|\textbf{W}\|^2_2+\beta\Gamma(a).
\end{align}
Here, $\mathcal{L}_{\rm{id}}/\mathcal{L}_{\rm{bd}}$ and $\mathcal{L}_{\rm{r}}(\mathcal{D}_{\rm{c}};\bm{\theta,\Lambda})$ respectively represent the initial/boundary data loss and residual data loss, they can be defined as following
\begin{align}\label{E7}
&\mathcal{L}_{\rm{id}}(\mathcal{D}_{\rm{i}};\bm{\theta})=\frac{1}{N^k_{\rm{i}}}\big\|\bm{q}^{\bm{\theta},\mathrm{i}}-\bm{q}^{\bm{m},\mathrm{i}}\big\|^2_2,\\
\label{E8}
&\mathcal{L}_{\rm{bd}}(\mathcal{D}_{\rm{b}};\bm{\theta})=\frac{1}{N^k_{\rm{b}}}\big\|\bm{q}^{\bm{\theta},\mathrm{b}}-\bm{q}^{\bm{m},\mathrm{b}}\big\|^2_2+\frac{1}{N^k_{\rm{b}}}\Big[\big\|\bm{q}^{\bm{\theta},\mathrm{lb}}-\bm{q}^{\bm{\theta},\mathrm{ub}}\big\|^2_2+\big\|\bm{q}_x^{\bm{\theta},\mathrm{lb}}-\bm{q}_x^{\bm{\theta},\mathrm{ub}}\big\|^2_2\Big],
\end{align}
and
\begin{align}\label{E9}
\mathcal{L}_{\rm{r}}(\mathcal{D}_{\rm{c}};\bm{\theta,\Lambda})=\frac{1}{N^k_{\mathrm{c}}}\big\|f^c_{\bm{q}}\big\|^2_2,
\end{align}
where $\|\cdot\|_2$ denotes the $L^2$ norm, $\bm{q}^{\bm{\theta},\mathrm{i}}$ and $\bm{q}^{\bm{\theta},\mathrm{b}}$ represent the learning results of $\bm{q}^{\bm{\theta}}$ acting on initial points $\mathcal{D}_{\mathrm{i}}=\{I_k\}$ and boundary points $\mathcal{D}_{\mathrm{b}}=\{B_k\}$. Here $\bm{q}^{\bm{\theta},\mathrm{b}}$ includes lower boundary $\bm{q}^{\bm{\theta},\mathrm{lb}}$ and upper boundary $\bm{q}^{\bm{\theta},\mathrm{ub}}$, the second item on the right of Eq. \eqref{E8} indicates that $\bm{q}^{\bm{\theta}}$ satisfies the periodic boundary condition. Besides, $\bm{q}^{\bm{m},\mathrm{i}}$ and $\bm{q}^{\bm{m},\mathrm{b}}$ represent the measurement data of $\bm{q}$ on initial points $\mathcal{D}_{\mathrm{i}}=\{I_k\}$ and boundary points $\mathcal{D}_{\mathrm{b}}=\{B_k\}$. The $f^c_{\bm{q}}$ is value of $f_{\bm{q}}$ on residual collocation points $\mathcal{D}_{\mathrm{c}}=\{C_k\}$.

Moreover, $\|\bm{W}\|_2$ drives the weights closer to the origin, and $\alpha$ is the regularization parameter. Then the $\Gamma(a)$ with weight parameter $\beta$ is the slope recovery term expressed as
\begin{align}\label{E10}
\Gamma(a)=\frac{1}{\frac{1}{D-1}\sum\limits_{d=1}^{D-1}\mathrm{exp}\Big(\frac{\sum\limits_{i=1}^{N_d}a_i^d}{N_d}\Big)}.
\end{align}
Here, the main reason behind including slope recovery term is that such term contributes to the gradient of the loss function without vanishing. The overall effect of adding this term is that it forces the network to increase the value of activation slope quickly thereby speeding up training process of PINN.

The resulting optimization problem demand to find the minimum value of total loss function by optimizing parameters, that is leads to finding the minimum value of the loss function by optimizing the parameters $\{\bm{\theta}\}$, that is, we seek
\begin{align}
&\{\bm{\theta}^*\}:=\mathop{\mathrm{arg\,min}}\limits_{\bm{\theta}\in\mathcal{N}}\mathcal{L}(\mathcal{D}_{\rm{i}},\mathcal{D}_{\rm{b}},\mathcal{D}_{\rm{c}};\bm{\theta,\Lambda}), \nonumber
\end{align}
where $\{\bm{\theta}^*\}$ mean the optimal set of parameters. In this work, we utilize Adam optimizer with continuous exponential decay learning rate to optimize the total loss function in each stage PINN \cite{Kingma-aX2014}, in which initial learning rate is 1e-3, decay velocity is 1000, and decay coefficient is 0.9.

\begin{figure}[tb]
\centering
\begin{minipage}[t]{0.99\textwidth}
\centering
\includegraphics[height=9cm,width=16cm]{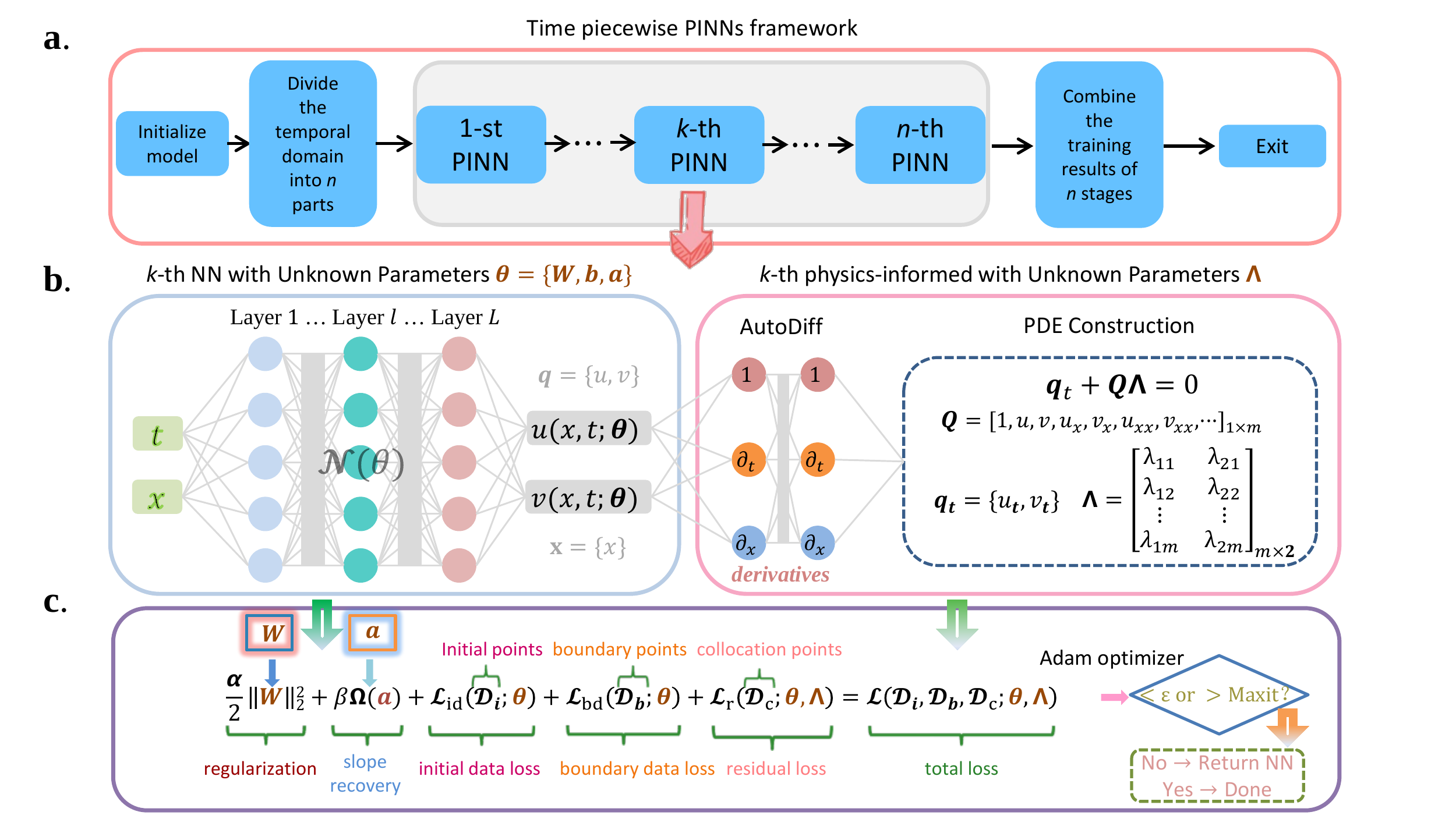}
\end{minipage}
\centering
\caption{(Color online) Schematic architecture of the tpPINNs model for data-driven forward and inverse problems of nonlinear systems. $\bm{\mathrm{a}}$ overall framework of tpPINNs. $\bm{\mathrm{b}}$ the NN part (left panel) and physics-informed part (right panel) of the $k$-th PINN in tpPINNs, The left NN is the universal approximation network without informed while the right network is physics-informed network dominated by the PDE construction, in which $\bm{q}=\{u,v\}$, $\bm{\mathrm{x}}=\{x\}$. $\bm{\mathrm{c}}$ the loss function and optimize process (Adam optimizer) of $k$-th PINN in tpPINNs, where the training parameters of total loss $\mathcal{L}$ are provided from NN part and physics-informed part.}
\label{F2}
\end{figure}

Once the training has completed the $n$ stages PINNs, we splice the training data of each subnet into the whole time region in time sequence. The Fig. \ref{F2} depicts the schematic architecture of the tpPINNs model, which is designed to tackle data-driven problems related to nonlinear systems. Fig. \ref{F2} $\bm{\mathrm{a}}$ exhibits the overall framework of tpPINNs, the tpPINNs algorithm that runs within the network involves several steps, including the initialization of the network model, the segmentation of the long time domain into $n$ stages, the successive running of $n$ stages of PINNs, the storage and combination of the training results for each stage, and the final termination of the algorithm program. Fig. \ref{F2} $\bm{\mathrm{b}}$ displays the network structure diagram of the $k$-th PINN stemming from Fig. \ref{F2} $\bm{\mathrm{a}}$, which including the NN part [left panel] and physics-informed part [right panel] of the $k$-th PINN in tpPINNs. The loss function and optimize process [Adam optimizer] of $k$-th
PINN are showcase in Fig. \ref{F2} $\bm{\mathrm{c}}$.

\subsection{Model system}
In binary BECs, the GPE with the LHY correction can describe QD. Especially, we consider that the binary BECs has symmetric spinor components, then require that the coupling constants describing the repulsion between each atom are equal, i.e. $\textsl{g}_{\uparrow\uparrow}=\textsl{g}_{\downarrow\downarrow}\equiv\textsl{g}$, and demand the numbers of atoms in the components are also equal. This results in the the equilibrium densities of both components are identical, which makes the analysis easier and the results clearer\cite{Astrakharchik-PRA2018}. In this case, the dynamics of one-dimensional QD can be described by the underlying time-dependent GPE \cite{Petrov-PRL2016,Astrakharchik-PRA2018,Lv-PLA2022}, as express in following
\begin{align}\label{MS-GPE}
\mathrm{i}\hbar\Psi_t=-\frac{\hbar^2}{2m}\Psi_{xx}+\delta\textsl{g}|\Psi|^2\Psi-\frac{\sqrt{2m}}{\pi\hbar}\textsl{g}^{3/2}|\Psi|\Psi,
\end{align}
here, $\hbar$ is quantum constant. In the two spinor components, the $\delta\textsl{g}=\textsl{g}_{\uparrow\downarrow}+\sqrt{\textsl{g}_{\uparrow\uparrow}\textsl{g}_{\downarrow\downarrow}}$ is positive coupling constant representing the mean field residual interaction, and $\textsl{g}=\sqrt{\textsl{g}_{\uparrow\uparrow}\textsl{g}_{\downarrow\downarrow}}$ is the positive coupling constant representing the intraspecies interaction. The difference $\delta\textsl{g}$ between attractive intercomponent and repulsive intracomponent interactions is responsible for appearance of a soft ``density'' mode, while the coupling constant $\textsl{g}$ is relevant for inducing a hard ``spin'' mode, and condition $\delta\textsl{g}\ll\textsl{g}$ causes a separation of scales. Experimentally, the value of $\delta\textsl{g}$ can be tuned to positive or negative values by Feshbach resonance\cite{Petrov-PRL2016,Astrakharchik-PRA2018,Lv-PLA2022}.

On the right side of Eq. \eqref{MS-GPE}, the first term indicates the kinetic energy term, the second term manifests the contact interaction, and the third term represents the LHY term. In our model, a superfluid state (QDs) form due to the mean field repulsion ($\delta\textsl{g}>0$) originates from the contact (local) intracomponent interaction and the quantum fluctuation attraction ($\textsl{g}>0$) provided by LHY term. We define following transformation,
\begin{align}\nonumber
x=\sqrt{\frac{\pi^2\delta\hbar^4}{2m^2\textsl{g}_0^2}}x',\,t=\frac{\pi^2\delta\hbar^3}{2m\textsl{g}_0^2}t',\,\Psi=\frac{\sqrt{2m\textsl{g}_0}}{\pi\delta\hbar}\psi,\,\textsl{g}=\textsl{g}_0G,
\end{align}
then the 1D model \eqref{MS-GPE} is transformed into a dimensionless one. For convenience, here we severally replace the labels $x'$ and $t'$ by using $x$ and $t$, and the final dimensionless model expresses as in following
\begin{align}\label{MS-GPE-dm}
\mathrm{i}\psi_t+\frac{1}{2}\psi_{xx}-G|\psi|^2\psi+G^{3/2}|\psi|\psi=0,
\end{align}
where $-G$ and $G^{3/2}$ represent the coefficient of the residual repulsive interaction from the mean field and the coefficient of attractive interaction from the LHY term, respectively. Specifically, if we set $\psi=u+\mathrm{i}v$, then from Fig. \ref{F2}, we can obtain
\begin{align}\nonumber
&\bm{Q}=[u_{xx},v_{xx},u^3,v^3,uv^2,vu^2,u\sqrt{u^2+v^2},v\sqrt{u^2+v^2}],\\ \nonumber
&\bm{\Lambda}=\begin{bmatrix} 0 & \frac12 & 0 & -G & 0 & -G & 0 & G^{3/2} \\ -\frac12 & 0 & G & 0 & G & 0 & -G^{3/2} & 0 \end{bmatrix}^{\rm{T}},\nonumber
\end{align}
where $\bm{\Lambda}$ is a sparse coefficient matrix. When $G=1$, the ground-state solution of Eq.\eqref{MS-GPE-dm} for the droplet can be found in an explicit form \cite{Petrov-PRL2016,Astrakharchik-PRA2018}:
\begin{align}\label{GPE-es}
\psi_{dp}(x,t)=-\frac{3\mu\mathrm{e}^{-\mathrm{i}\mu t}}{1+\sqrt{1+\frac92\mu}\cosh\big(\sqrt{-2\mu x^2}\big)}.
\end{align}
The relation between chemical potential $\mu$ and normalization $N$ is revealed by
\begin{align}\label{CP-NA}
N=\frac43\Bigg[\ln\Bigg(\frac{1+\sqrt{-\frac92\mu}}{\sqrt{1+\frac92\mu}}\Bigg)-\sqrt{-\frac92\mu}\Bigg].
\end{align}
When the number of atoms exceeds a certain threshold, the one-dimensional QDs will form a flat top. With $N$ increasing, the QDs get wider and wider. Different from the general single-soliton solution of the one-dimensional NLS with attractive cubic nonlinearity \cite{Zakharov-ZETF1972}, our droplet has a flat bulk region for $N\rightarrow\infty, \mu\rightarrow-2/9$, namely Eq.\eqref{GPE-es} becomes the following plane wave $\psi_{dp}(x,t)=2/3\mathrm{e}^{\frac29\mathrm{i}t}$, in which the amplitude is 2/3.

In this work, for overcoming the challenge of long time domain during model training, we propose the tpPINNs for studying complex dynamics of QD in BECs, then we focus on complex dynamics of QDs in a long time region for the model \eqref{MS-GPE-dm} from initial-boundary value data by means of tpPINNs. We select $n=10,a_i^d=0.1$ as the initialization of scalable parameters in the adaptive activation function, and choose the hyperbolic tangent ($\tanh$) as activation function. Furthermore, we select the multi-layer perceptrons with the Xavier initialization in NN parts. The reference dynamics and initial/boundary points are obtained by means of ``Chebfun-numerical computing with functions'' method, and the residual collocation points for training are generated by the Latin Hypercube Sampling method (LHS) \cite{Stein-T1987}.

\section{Computational examples}
In this section, we consider the complex dynamics on the one-dimensional QDs by using tpPINNs, including the intrinsic modulations of a single droplets, collisions between droplets and breather excitations on droplet background.

\subsection{Intrinsic modulations in a single droplet}
In this subsection, we study density-modulation excitation modes with a certain wave number in a single droplet, and address their stability. The ``breathing'' or monopole mode is the lowest compression mode, with the spinor components moving in-phase, making the droplet size periodically oscillating. We excite this mode by driving a single droplet out of equilibrium and study the ensuing dynamics, simulating Eq. \eqref{MS-GPE-dm} with $G=1$ via tpPINNs. The excitation of internal dynamics of the droplet is provided by imposing periodic density modulation with wave number $\kappa$ onto it, thus corresponding initial wave function [initial condition $\mathcal{I}$] is shown as
\begin{align}\label{SD-iwf1-ic}
\psi(x,t=0)=\psi_{dp}(x)\cos(\kappa x),\,x\in[x_{\mathrm{lb}},x_{\mathrm{ub}}],
\end{align}
in which $\psi_{dp}(x)$ is the exact form of the ground-state solution \eqref{GPE-es} at initial moment $t=0$. For small droplets with small $N$ the dependence of the chemical potential has a power-law form \cite{Astrakharchik-PRA2018}, here we select $\mu=-0.379N^{2/3}$. Furthermore, we consider following mixed boundary conditions $\mathcal{B}$
\begin{align}\label{SD-iwf1-bc}
\psi(x_{\mathrm{lb}},t)=\psi(x_{\mathrm{ub}},t),\,\psi_x(x_{\mathrm{lb}},t)=\psi_x(x_{\mathrm{ub}},t),\,\psi(x_{\mathrm{b}},t)=\psi^{\bm{m},\mathrm{b}},\,t\in[0,T].
\end{align}
As a result, in direct simulations the perturbed droplet may keep its shape entirety or suffer fragmentation, depending on $N$ and $\kappa$.

$\bullet$ droplet avoiding fragmentation [$N=0.46, \kappa=0.15$]

For avoiding the fragmentation of the droplet, we need to the energy of excitation is much smaller than the potential barrier induced by the surface tension, at this time, we select $N=0.46, \kappa=0.15$. Then, we consider the initial-boundary conditions \eqref{SD-iwf1-ic}-\eqref{SD-iwf1-bc} in the spatiotemporal region $[0,150]\times[-30,30]$. Here, one can observe the quite large time domain from 0 to 150, this is a significant challenge for studying data-driven problems via conventional PINN. Thus, we use 15 subnet ($0\leqslant k\leqslant15$) in tpPINNs, namely 15 stages sub-PINNs. In particular, we choose a time interval length of 10 for each stage by evenly dividing the whole time domain. In each stage, we utilize the 5 hidden-layer NNs with 200 neurons per hidden layer to reveal the dynamic behavior of droplet to avoid fragmentation by using Adam optimizer to minimize the total loss \eqref{E6}. Furthermore, the initial/boundary training points $\mathcal{D}_{\rm{i}}/\mathcal{D}_{\rm{b}}$, residual collocation points $\mathcal{D}_{\rm{c}}$, penalty coefficient $\alpha$ of regularization term and weight $\beta$ of slope recovery term is arbitrarily select at each stage. The training information of tpPINNs with 15 subnet is shown in table \ref{Tab1} [see Appendix \ref{App-A}], in which including network hidden-layers, initial/boundary training points, residual collocation points, hyper-parameters, Adam optimization and corresponding $L^2$ norm error.

\begin{figure}[htbp]
\centering
\subfigure[]{
\begin{minipage}[t]{0.48\textwidth}
\centering
\includegraphics[height=4.5cm,width=7cm]{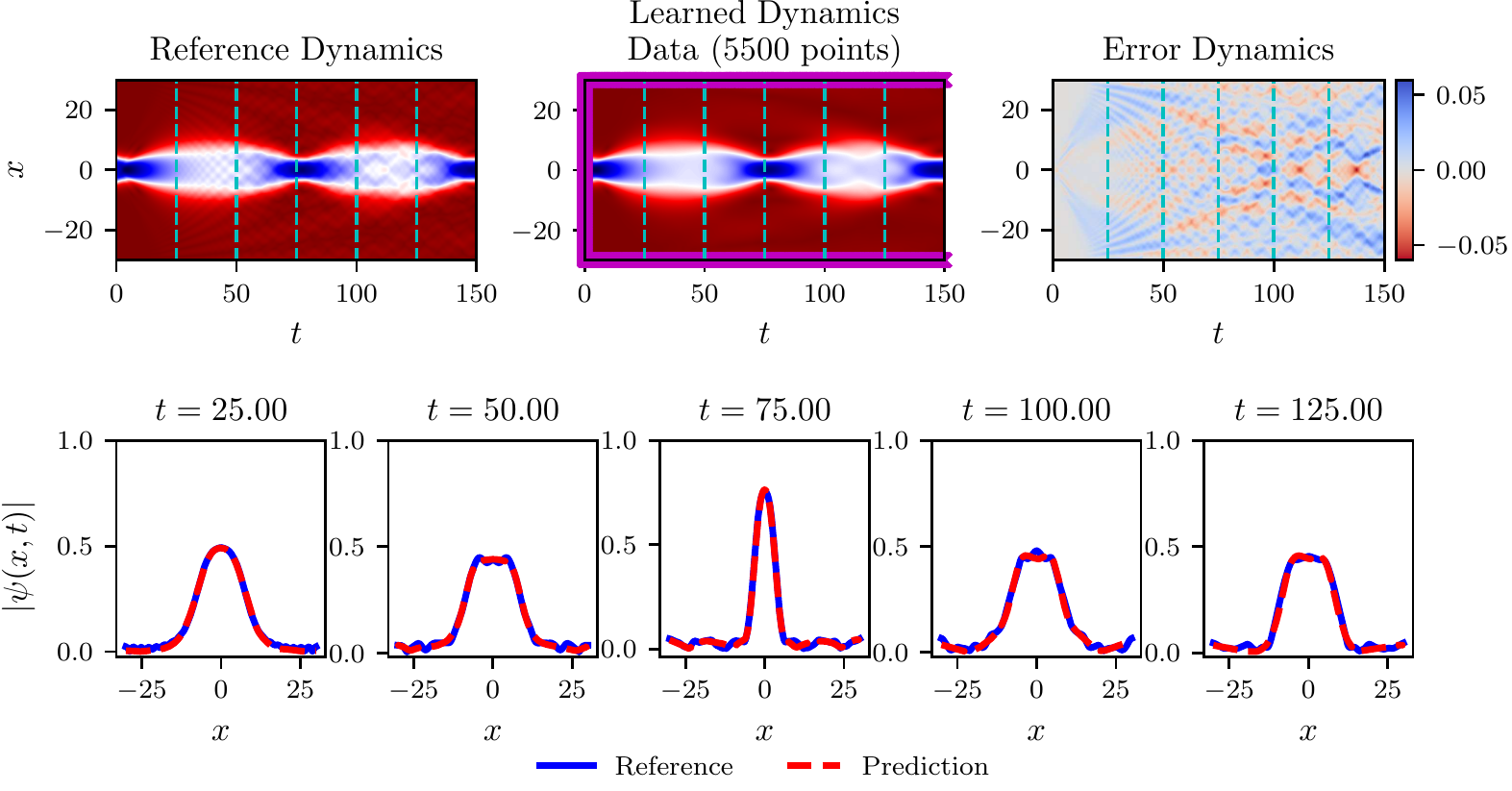}
\end{minipage}
}%
\subfigure[]{
\begin{minipage}[t]{0.48\textwidth}
\centering
\includegraphics[height=5cm,width=7cm]{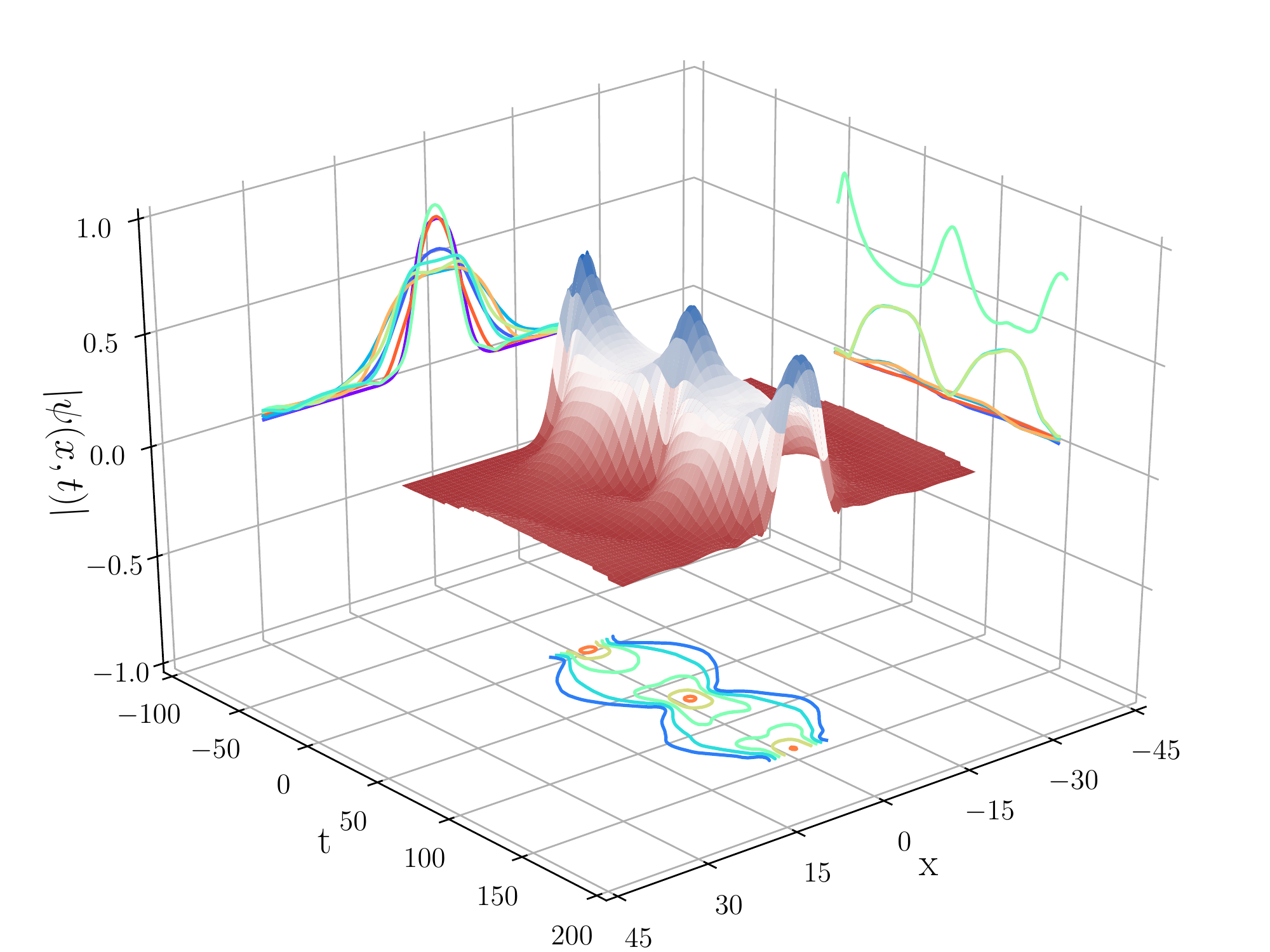}
\end{minipage}%
}%
\centering
\caption{(Color online) The training results of single droplet $\psi$, which it avoids fragmentation, arising from the tpPINNs with 5500 initial and boundary points (mediumorchid $``\times"$ in learned dynamics): (a) The reference, learned and error dynamics density plots, as well as sectional drawings which contain the learned and reference single droplet at five distinct moments $t=25, 50, 75, 100$ and 125 (darkturquoise dashed lines in density plots); (b) The 3D plot with contour map for the data-driven single droplet.}
\label{F3}
\end{figure}

By randomly providing 5500 initial and boundary points from the initial-boundary value conditions, the relative $L^2$ norm error of the tpPINNs model achieves 5.362393$\rm e$-02 for single droplet $\psi$ [see Tab. \ref{Tab1}]. Fig. \ref{F3} manifests the deep learning results of the data-driven single droplet $\psi$ stemming from the tpPINNs for the GPE \eqref{MS-GPE-dm}. In Fig. \ref{F3}(a), we display the density plots of the reference dynamics, learned dynamics and error dynamics, then showcase its corresponding amplitude scale size on the right side of density plots, and exhibit the sectional drawings which contain the learned and reference droplet at five different moments. Fig. \ref{F3}(b) indicates the 3D plot with contour map on three planes for the predicted single droplet. In this case, we observed almost periodic oscillations in the width of the droplet. The evolution curve figures of the loss function arising from the tpPINNs with 15 stage subnets are displayed in Fig \ref{F-loss1} [see Appendix \ref{App-B}].

$\bullet$ droplet splitting [$N=0.46, \kappa=1$]

In this part, for exploring more interesting dynamic behavior of single droplet, we increase the value of $\kappa$ [especially $\kappa=1$], then the tpPINNs and the initial-boundary conditions \eqref{SD-iwf1-ic}-\eqref{SD-iwf1-bc} of Gross-Pitaevskii equation \ref{MS-GPE-dm} are used to train the different dynamic behavior of the droplets in the spatiotemporal region $[0,25]\times[-30,30]$. Then we averagely divide the temporal region $[0,25]$ into 5 sub-domains, thus obtain 5 sub-PINNs with same time interval length in tpPINNs [here $k=5$]. Accordingly, the training information of each stage is exhibited in the table \ref{Tab2} [see Appendix \ref{App-A}].

\begin{figure}[htbp]
\centering
\subfigure[]{
\begin{minipage}[t]{0.48\textwidth}
\centering
\includegraphics[height=4.5cm,width=7cm]{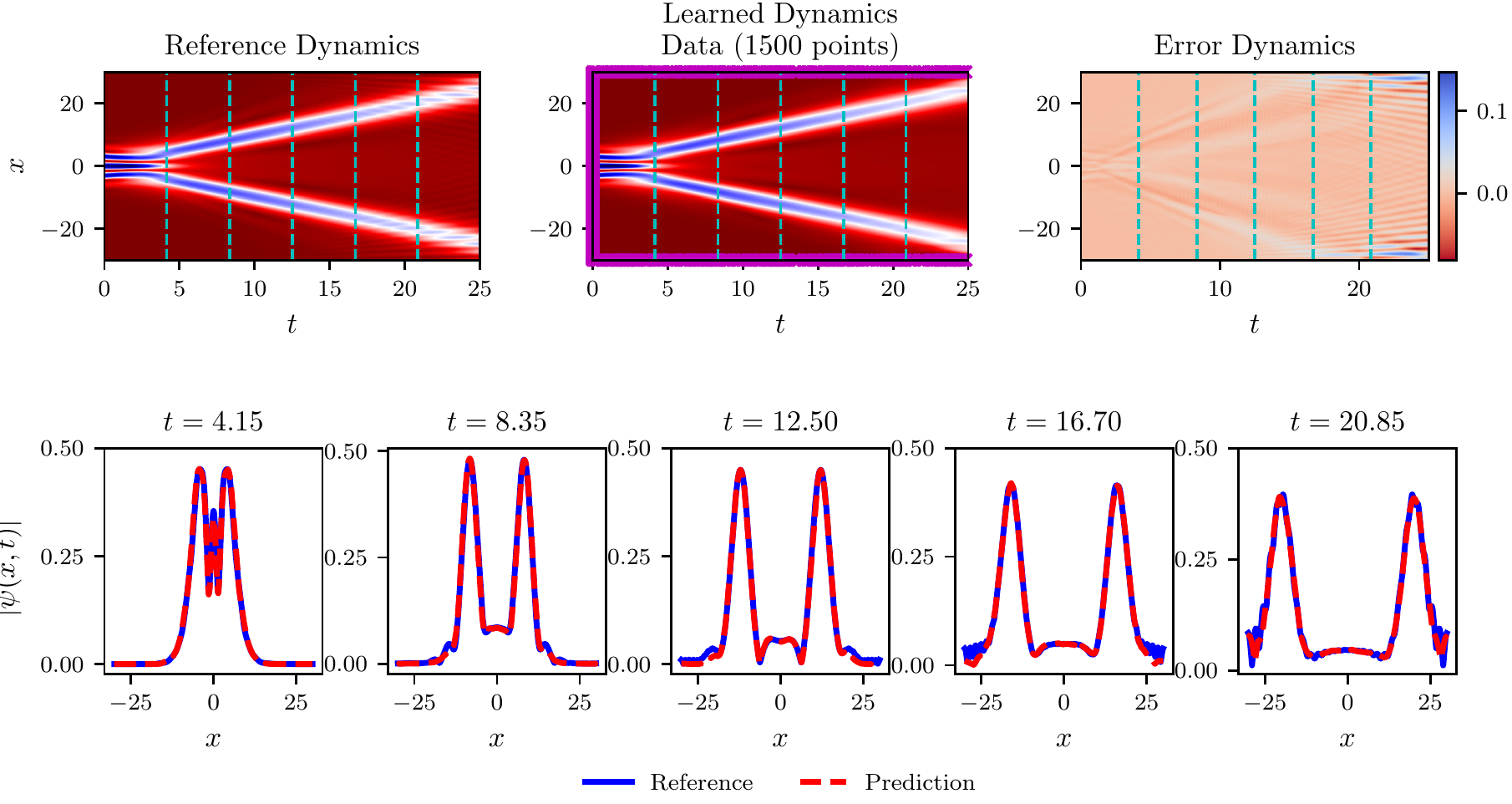}
\end{minipage}
}%
\subfigure[]{
\begin{minipage}[t]{0.48\textwidth}
\centering
\includegraphics[height=5cm,width=7cm]{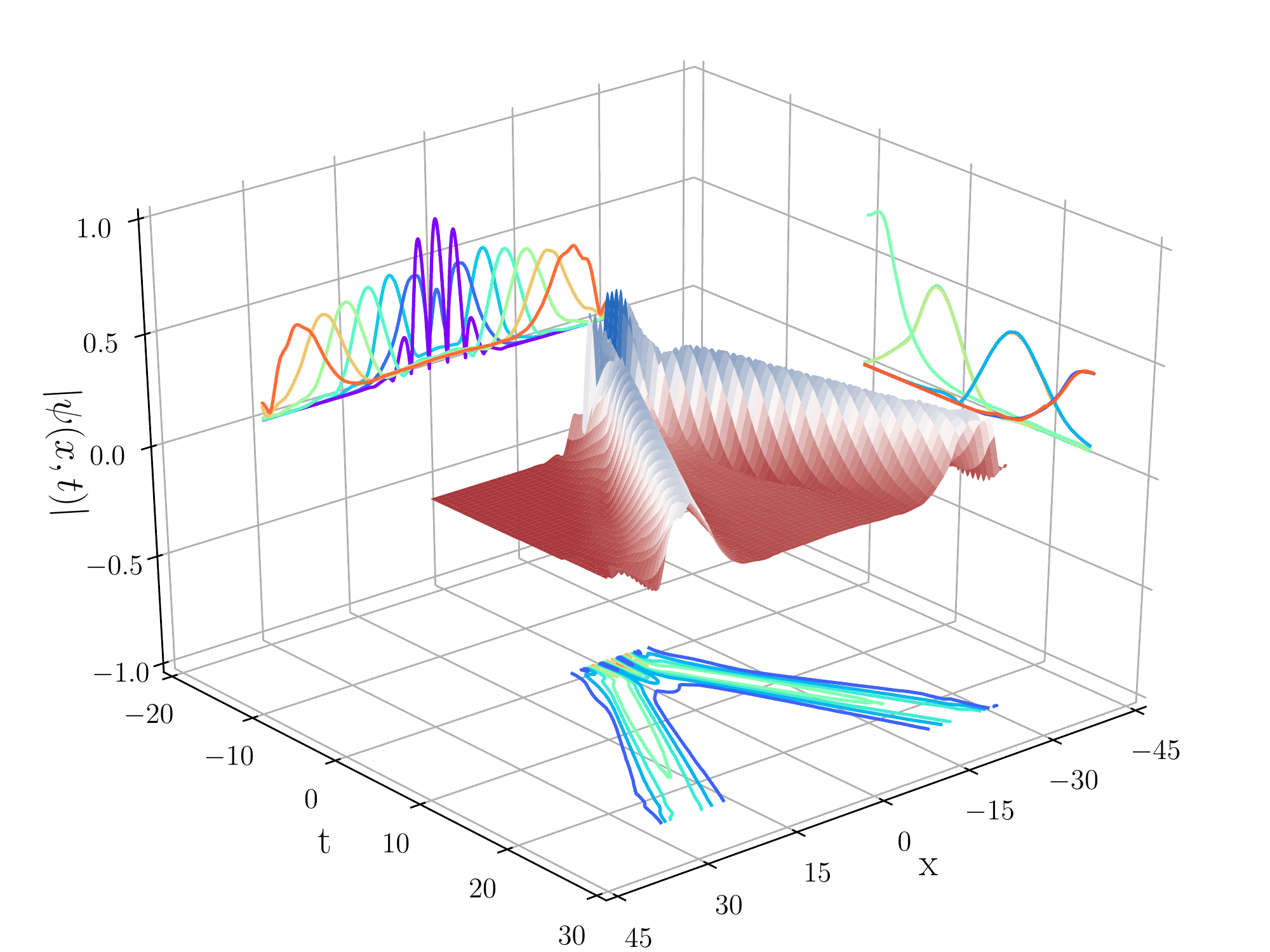}
\end{minipage}%
}%
\centering
\caption{(Color online) The training results of single splitting droplet $\psi$ arising from the tpPINNs with 1500 initial and boundary points (mediumorchid $``\times"$ in learned dynamics): (a) The reference, learned dynamics and error dynamics density plots, as well as sectional drawings which contain the learned and reference single droplet $\psi$ at five distinct moments $t=4.15, 8.35, 12.50, 16.70$ and 20.85 (darkturquoise dashed lines in density plots); (b) The 3D plot with contour map for the data-driven single splitting droplet.}
\label{F4}
\end{figure}

We use 1500 initial and boundary points in tpPINNs, then obtain the 6.993923$\rm e$-02 relative $L^2$ norm error for data-driven splitting droplet $\psi$ in the tpPINNs framework [see Tab. \ref{Tab2}]. Fig. \ref{F4} displays the data-driven training results of the splitting droplet $\psi(x,t)$ by utilizing the tpPINNs with the initial-boundary value conditions of the GPE. The upper panel of Fig. \ref{F4}(a) depicts various dynamic density plots, including reference, learned dynamics as well as error dynamics with corresponding amplitude scale size on the right side, and the bottom panel of Fig. \ref{F4}(a) presents sectional drawing at different moments. The 3D plot with contour map on three planes for the data-driven splitting droplet has been displayed in Fig. \ref{F4}(b). One can observe that the single droplet splits in two or more escaping fragments (which are smaller droplets) which fly away, as shown in Fig. \ref{F4}. Indeed, there has a critical value $\kappa_c$, once the large $\kappa$ exceeds this value, the droplet splits into several fragments, they never recombine \cite{Astrakharchik-PRA2018}. The evolution curve figures of the loss function arising from the tpPINNs with 5 stage subnets are displayed in Fig \ref{F-loss2} [see Appendix \ref{App-B}].

To further showcase the effectiveness of tpPINNs in studying QD within the context of BECs, we provide a comparison of the training errors between tpPINNs and conventional PINNs in table \ref{Tab3.1-2}, which spans same long time domain. Here both network structures are provided equal number of initial/boundary points, although conventional PINN have a limitation on the number of residual collocation points due to computer memory, this is typically not a major concern for us. From this table, we can observe that the relative $L^2$ norm error of conventional PINN using only one single network is quite large when dealing with the long time region, while the training $L^2$ norm error of tpPINNs is more than 8 times smaller than that of traditional PINNs. In a word, tpPINNs exhibit superior learning performance and require less memory when training complex dynamic behaviors of QD over the long time domain.

\begin{table}[htbp]
  \caption{Error comparison between tpPINNs and conventional PINN in case of droplet splitting}
  \label{Tab3.1-2}
  \centering
  \scalebox{0.8}{
  \begin{tabular}{l|c|c|c|c|c|c|c|c}
  \toprule
  \diagbox{\small{\textbf{Networks}}}{\small{\textbf{Types}}} & $L,\,N_d$ & $t$ & $N^k_{\rm{i}}$ & $N^k_{\rm{b}}$ & $N^k_{\rm{c}}$ & $\alpha,\,\beta$ & Adam & $L^2$ error \\
  \hline
  PINN   & 5,\,200 & [0,25] & 1000 & 250 & 200000 & $0.00001,\,1$ & 250000 & 5.691071$\rm e$-01 \\
  \hline
  tpPINNs & \diagbox{}{} & [0,25] & 1000 & 250 & \diagbox{}{} & \diagbox{}{} & \diagbox{}{} & 6.993923$\rm e$-02 \\
  \bottomrule
  \end{tabular}}
\end{table}

\subsection{Collision between two droplets}
As is known to all, the soliton, which can maintain its shape while moving at a constant velocity, is the nonuniform coherent wave self-trapped owing to the action of the nonlinear dispersion. In our case, the wave function $\psi(x,t)$ \eqref{GPE-es} is formed by the interplay of quadratic and cubic terms in GPE \eqref{MS-GPE-dm} for the droplet, similarly to the shape of the usual bright soliton. Furthermore, we find that another noteworthy feature of soliton in (nearly) integrable systems is that their shape cannot be changed as they collide with another soliton. Thus, verifying a possible persistence for the droplets shape is crucial in the case of the pairwise collisions. To address this issue in this section, we study the collisions between two droplets for GPE \eqref{MS-GPE-dm} by using the tpPINNs after providing initial and boundary data. We take the initial wave function [initial condition] as a set of two droplets, as shown in following
\begin{align}\label{TD-iwf2-ic}
\psi(x,t=0)=\mathrm{e}^{\mathrm{i}mx/2+\varphi}\psi_1(x+x_0)+\mathrm{e}^{-\mathrm{i}mx/2}\psi_2(x-x_0),\,x\in[x_{\mathrm{lb}},x_{\mathrm{ub}}],
\end{align}
where $¡Àm/2$ are initial momenta of the colliding droplets, and $\varphi$ is the relative phase. $\psi_1(x)$ and $\psi_2(x)$ are the stationary shapes of droplets with normalization $N_1$ and $N_2$, borrowed from Eq. \eqref{GPE-es} at $t=0$, while chemical potentials are taken as $\mu_{1,2}=-2/9+(8/9)\mathrm{exp}(-2-(3/2)N_{1,2})$, $\pm x_0$ are their initial positions. Moreover, we still choose the boundary condition is \eqref{SD-iwf1-bc}. Then we employ tpPINNs to investigate three complex dynamic behaviors related to the collision of two droplets, each driven by distinct initial-boundary data.

$\bullet$ The first dynamic behaviors [$N_1=0.1,\,N_2=0.1,\,m=4,\,x_0=10,\,\varphi=0$]

In this part, we take $N_1=0.1,\,N_2=0.1,\,m=4,\,x_0=10,\,\varphi=0$ in the initial wave function \eqref{TD-iwf2-ic}, then study data-driven colliding droplets in spatiotemporal region $[0,10]\times[-30,30]$ by using a tpPINNs with $k=4$ subnets. The detailed training data of the network at each stage is shown in the table \ref{Tab3.2-1} [see Appendix \ref{App-A}].

\begin{figure}[tb]
\centering
\subfigure[]{
\begin{minipage}[t]{0.48\textwidth}
\centering
\includegraphics[height=4.5cm,width=7cm]{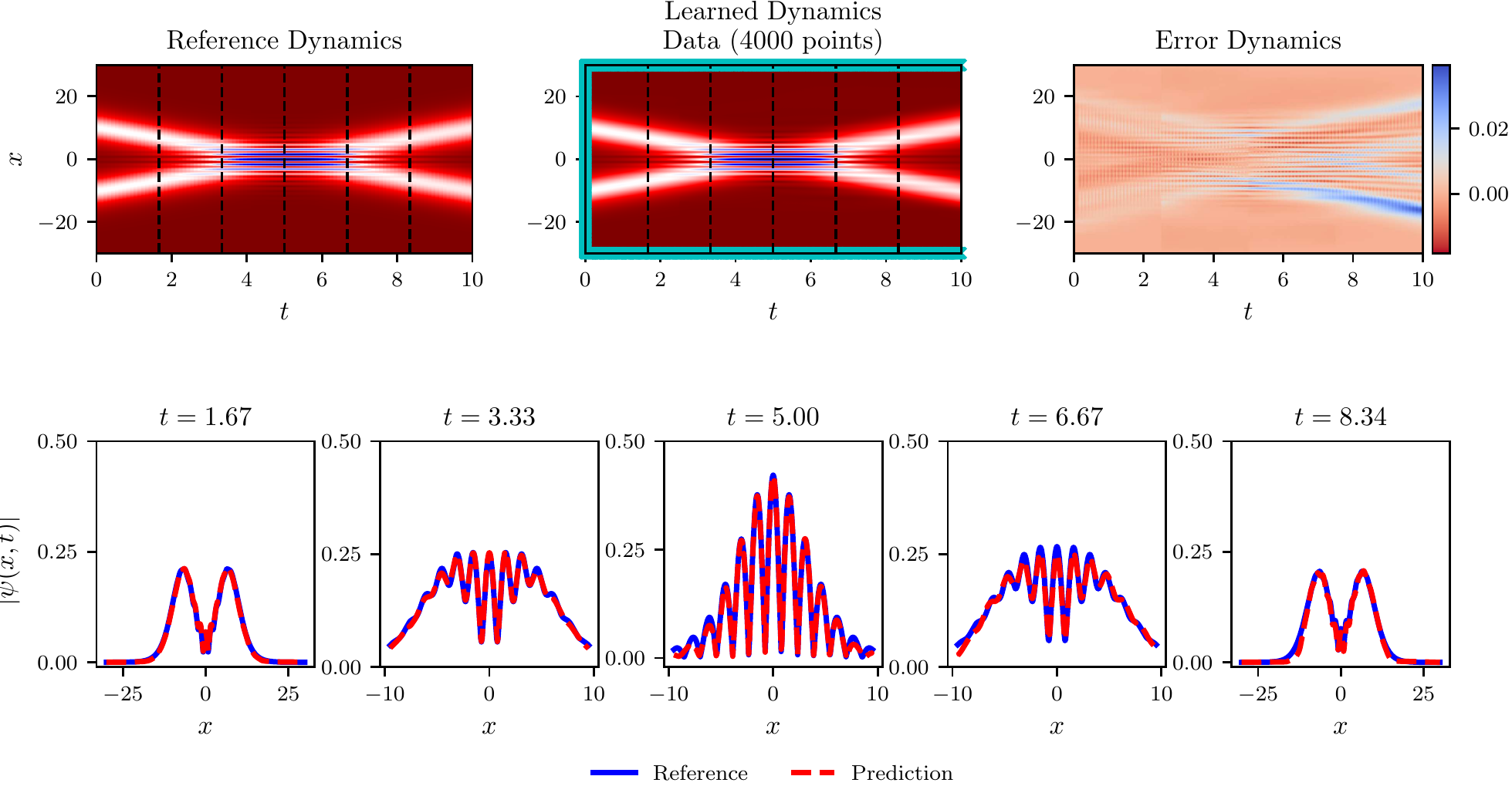}
\end{minipage}
}%
\subfigure[]{
\begin{minipage}[t]{0.48\textwidth}
\centering
\includegraphics[height=5cm,width=7cm]{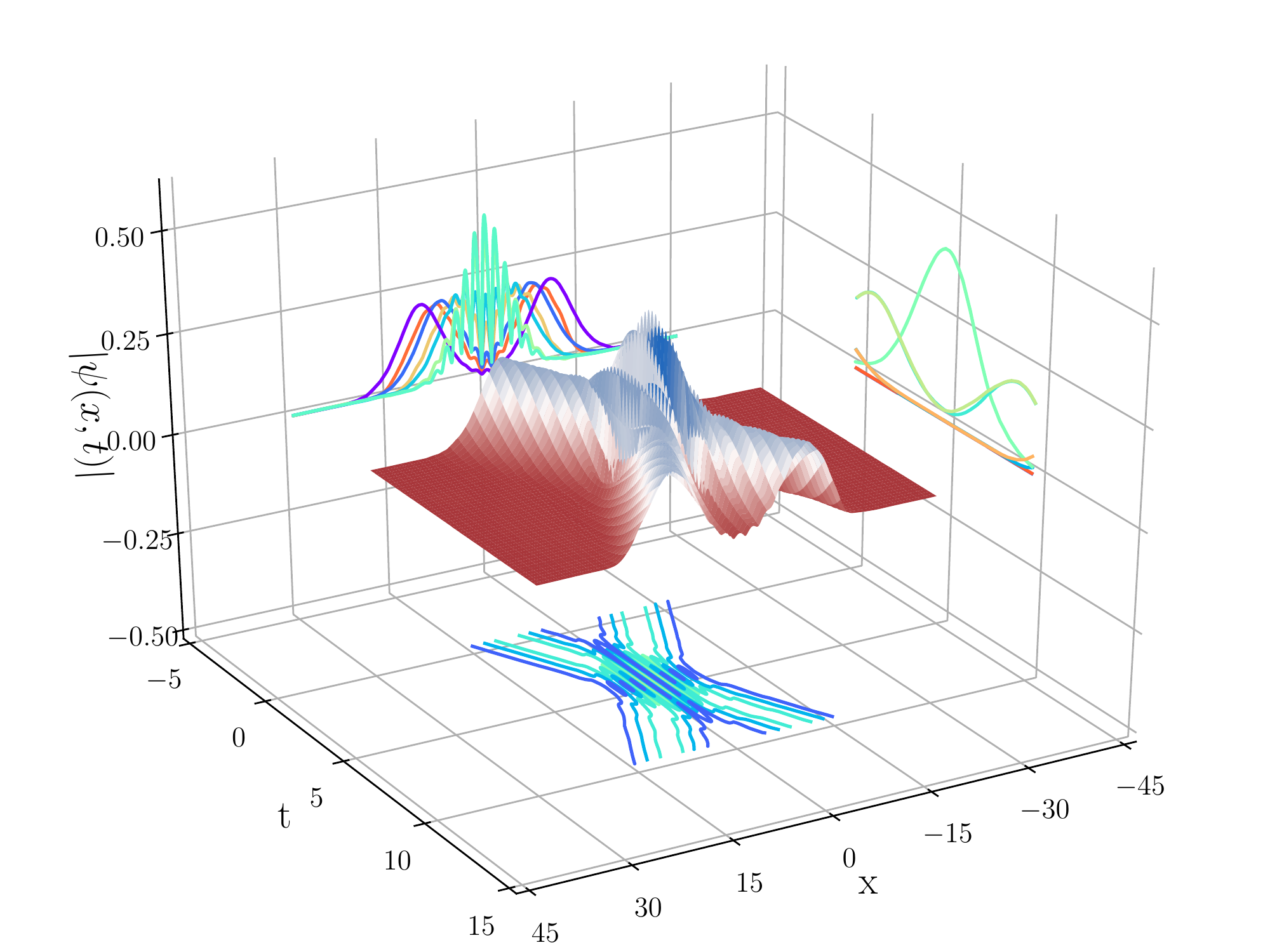}
\end{minipage}%
}%
\centering
\caption{(Color online) The training results for colliding droplets $\psi$ of first dynamic behaviors arising from the tpPINNs with 4000 initial and boundary points (cyan $``\times"$ in learned dynamics): (a) The reference, learned dynamics and error dynamics density plots, as well as sectional drawings which contain the learned and reference colliding droplet $\psi$ at five distinct moments $t=1.67, 3.33, 5.00, 6.67$ and 8.34 (black dashed lines in density plots); (b) The 3D plot with contour map for the data-driven colliding droplets.}
\label{F5}
\end{figure}

From table \ref{Tab3.2-1}, one can see that total 4000 initial and boundary points have been used for obtaining data-driven colliding droplets in tpPINNs, then relative $L^2$ norm error reaches 5.919966$\rm e$-02. Fig. \ref{F5} displays the training results of the colliding droplets arising from the tpPINNs. The upper panel of Fig. \ref{F5}(a) showcases three density plots with corresponding amplitude scale size on the right side, namely reference dynamics, learned dynamics as well as error dynamics, while sectional drawings at five moments are presented in the bottom panel of Fig. \ref{F5}(a). Owing to the quite dense dynamic behavior during the collision of two droplets, we shorten the spatial scale to $[-10,10]$ for observing more clearly the colliding dynamic behavior in the profile Fig. \ref{F5}(a) of the middle three moments. Fig. \ref{F5}(b) exhibits the 3D plot with contour map on three planes. Fig. \ref{F5} indicates that the shape of each droplet is precisely preserved after the collision. During the collision process [see Fig. \ref{F5}(a) about at time domain $[3,7]$], an interference pattern is well visible, which is a significant feature of the interplay of coherent matter waves. In the non collision phase, the dynamic behavior of the two droplets is similar to the two solitons of NLS. The evolution curve figures of the loss function arising from the tpPINNs with 4 stage subnets are displayed in Fig \ref{F-loss3} [see Appendix \ref{App-B}].

Similarly, we also exhibit that comparison of relative $L^2$ norm error between tpPINNs and conventional PINNs in table \ref{Tab3.2-2}. From this table, we observe that the relative $L^2$ norm error of conventional PINN using only one single network is quite large when studying the long time domain, while the training $L^2$ norm error of tpPINNs is more than 10 times smaller than that of conventional PINNs. This finding further indicates the advantages of tpPINNs in dealing with the intricate dynamics of QD over long time domains.

\begin{table}[htbp]
  \caption{Error comparison between tpPINNs and conventional PINN in first case of colliding droplets}
  \label{Tab3.2-2}
  \centering
  \scalebox{0.8}{
  \begin{tabular}{l|c|c|c|c|c|c|c|c}
  \toprule
  \diagbox{\small{\textbf{Networks}}}{\small{\textbf{Types}}} & $L,\,N_d$ & $t$ & $N^k_{\rm{i}}$ & $N^k_{\rm{b}}$ & $N^k_{\rm{c}}$ & $\alpha,\,\beta$ & Adam & $L^2$ error \\
  \hline
  PINN   & 5,\,200 & [0,10] & 800 & 1600 & 160000 & $0.0001,\,1$ & 200000 & 9.866230$\rm e$-01 \\
  \hline
  tpPINNs & \diagbox{}{} & [0,10] & 800 & 1600 & \diagbox{}{} & \diagbox{}{} & \diagbox{}{} & 5.919966$\rm e$-02 \\
  \bottomrule
  \end{tabular}}
\end{table}

$\bullet$ The second dynamic behaviors [$N_1=0.2,\,N_2=0.1,\,m=0.4,\,x_0=10,\,\varphi=0$]

To explore more interesting collision dynamics of QD, based on the first dynamic behaviors, we increase the value of normalization $N_1$ and decrease the value of initial momenta $m$, namely $N_1=0.2,\,N_2=0.1,\,m=0.4,\,x_0=10,\,\varphi=0$. Then, we employ tpPINNs with $k=9$ subnets to study the data-driven colliding droplets $\psi$, and the spatiotemporal domain is chosen as $[0,90]\times[-30,30]$. The corresponding training information is shown in the table \ref{Tab3.2-3} [see Appendix \ref{App-A}].

\begin{figure}[htbp]
\centering
\subfigure[]{
\begin{minipage}[t]{0.48\textwidth}
\centering
\includegraphics[height=4.5cm,width=7cm]{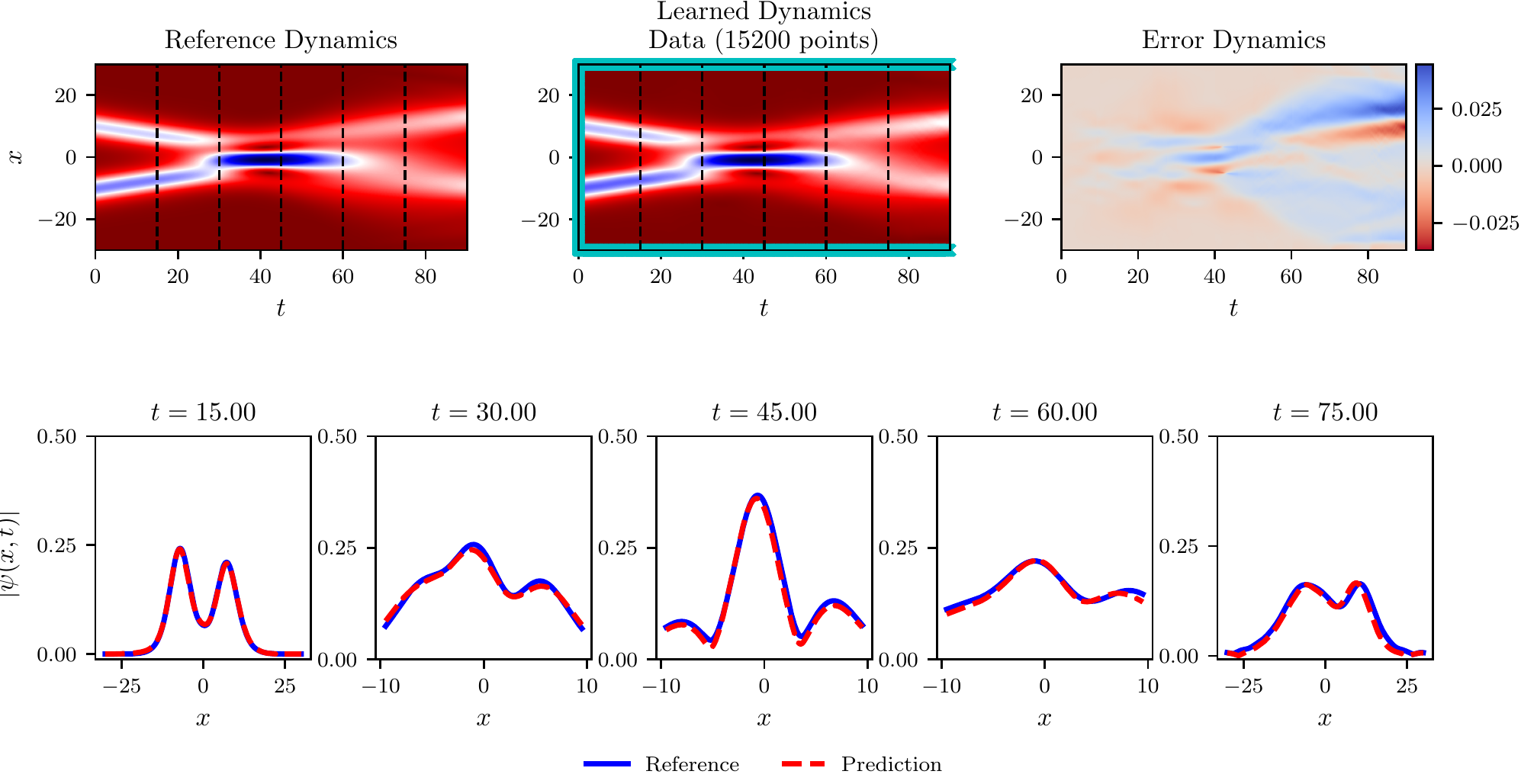}
\end{minipage}
}%
\subfigure[]{
\begin{minipage}[t]{0.48\textwidth}
\centering
\includegraphics[height=5cm,width=7cm]{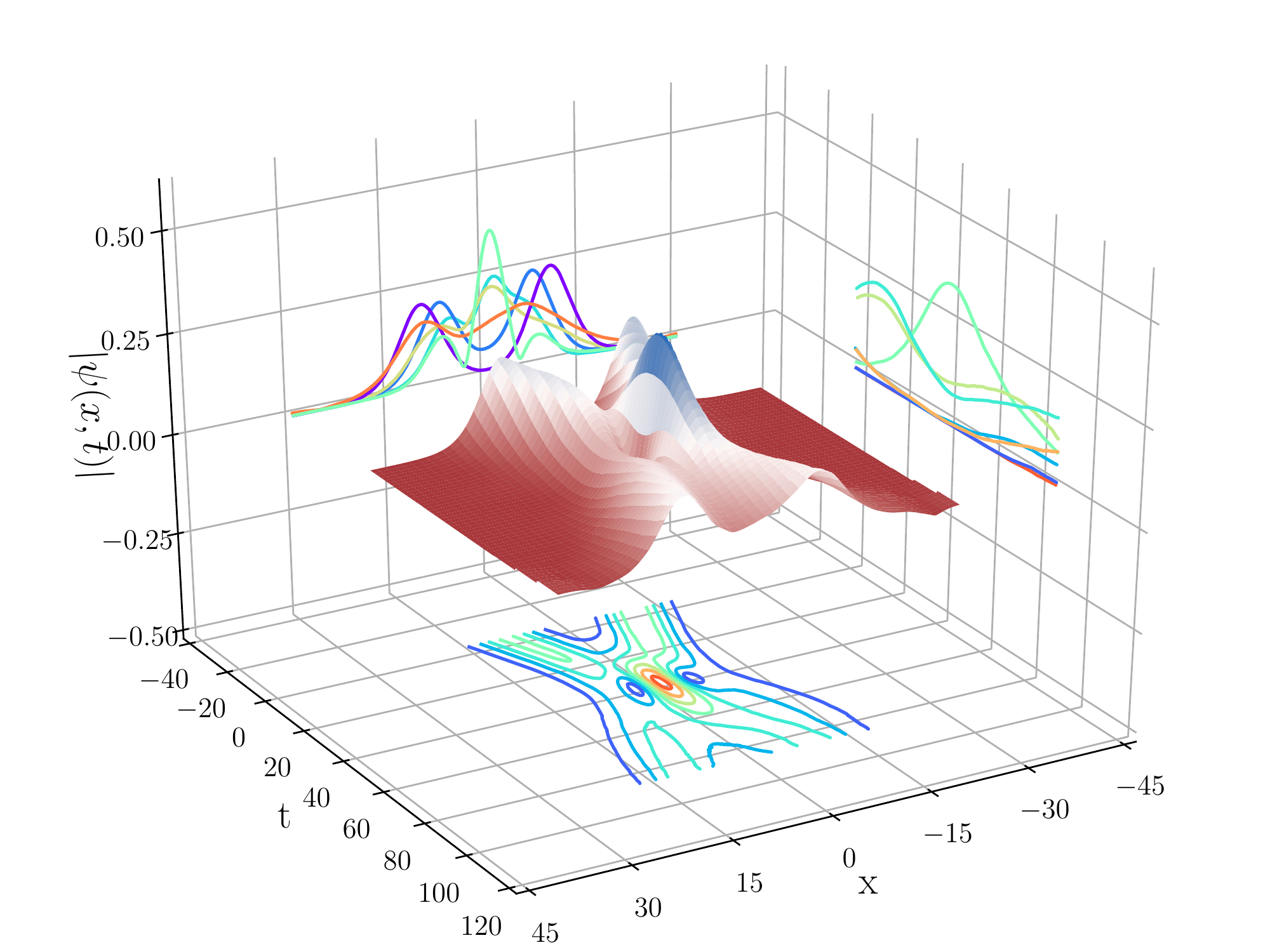}
\end{minipage}%
}%
\centering
\caption{(Color online) The training results for colliding droplets $\psi$ of second dynamic behaviors arising from the tpPINNs with 15200 initial and boundary points (cyan $``\times"$ in learned dynamics): (a) The reference, learned dynamics and error dynamics density plots, as well as sectional drawings which contain the learned and reference colliding droplet $\psi$ at five distinct moments $t=15, 30, 45, 60$ and 75 (black dashed lines in density plots); (b) The 3D plot with contour map for the data-driven colliding droplets.}
\label{F6}
\end{figure}

The relative $L^2$ norm error reaches 8.071953$\rm e$-02 result from the tpPINNs with using 15200 initial and boundary points. Fig. \ref{F6} displays the training results of the data-driven colliding droplets stemming from the tpPINNs. Similarly, the abundant density plots and sectional drawings are revealed in Fig. \ref{F6}(a), then the 3D plot and its contour map on three planes are exhibited in Fig. \ref{F6}(b). Compared with the first dynamic behaviors of colliding droplets, the colliding droplets in this part moves slowly, and the interference pattern does not appear during the collision process. Different from the training result for the two-soliton of NLS \cite{Pu-CPB2021}, from Fig. \ref{F6}, we have observed that the symmetry of droplet dynamics is broken at the left and right time points [$t=30$ and $t=60$], which are equidistant from the collision center [$t=45$]. Fig. \ref{F-loss4} [see Appendix \ref{App-B}] illustrates the evolution curve figures of the loss function resulting from tpPINNs with 9-stage subnets.

$\bullet$ The third dynamic behaviors [$N_1=0.2,\,N_2=0.1,\,m=0.4,\,x_0=10,\,\varphi=\pi/4$]

In this section, we focus on the effect of relative phase $\varphi$ on the colliding droplets $\psi$, thus we set $\varphi=\pi/4$ [all other parameters remain the same as those used for the second dynamic behaviors]. Then we consider the initial-boundary conditions \eqref{TD-iwf2-ic} and \eqref{SD-iwf1-bc} in the spatiotemporal region $[0,90]\times[-30,30]$. The corresponding training information of tpPINNs is shown as table \ref{Tab3.2-4} [see Appendix \ref{App-A}].

\begin{figure}[htbp]
\centering
\subfigure[]{
\begin{minipage}[t]{0.48\textwidth}
\centering
\includegraphics[height=4.5cm,width=7cm]{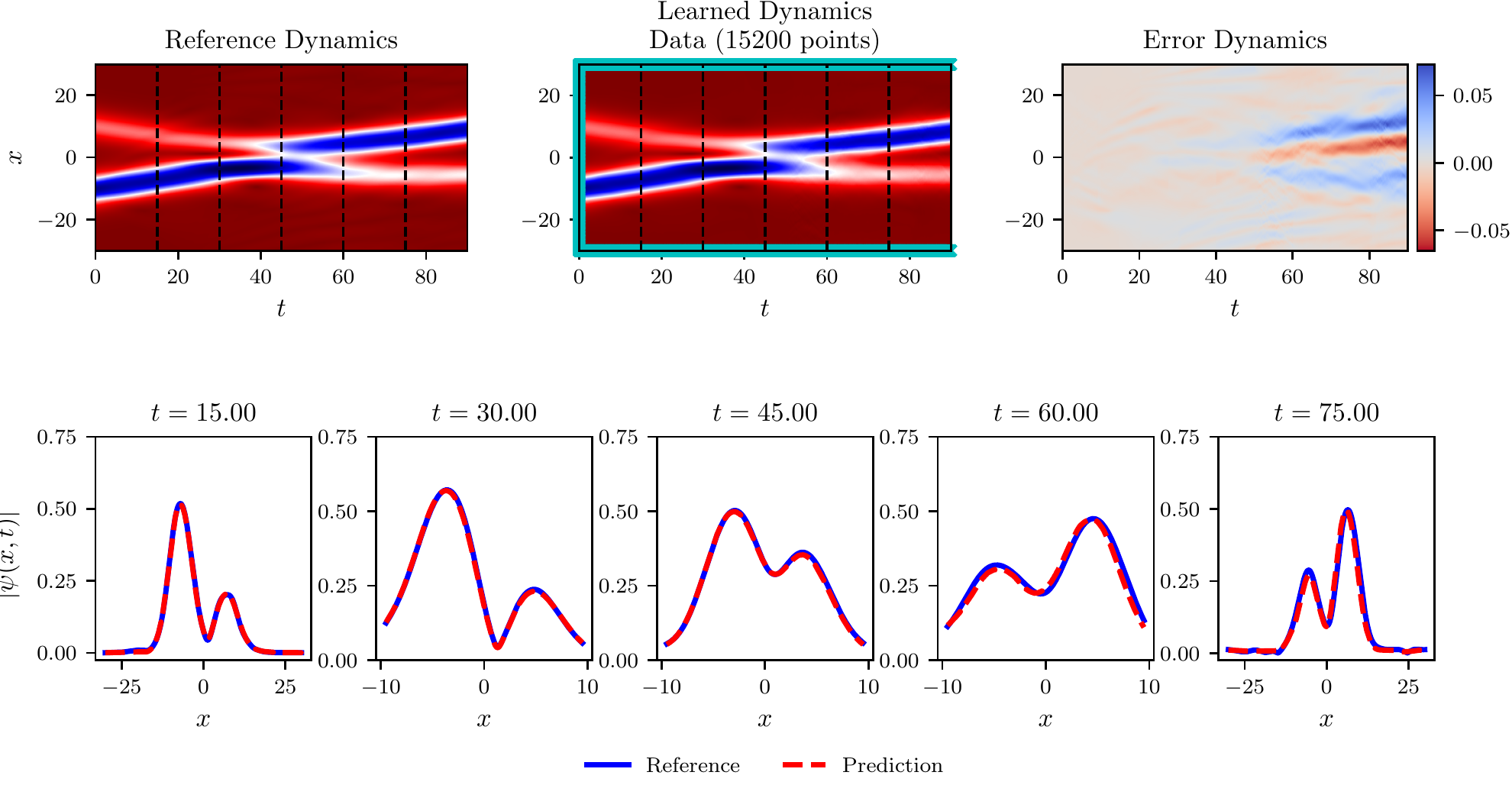}
\end{minipage}
}%
\subfigure[]{
\begin{minipage}[t]{0.48\textwidth}
\centering
\includegraphics[height=5cm,width=7cm]{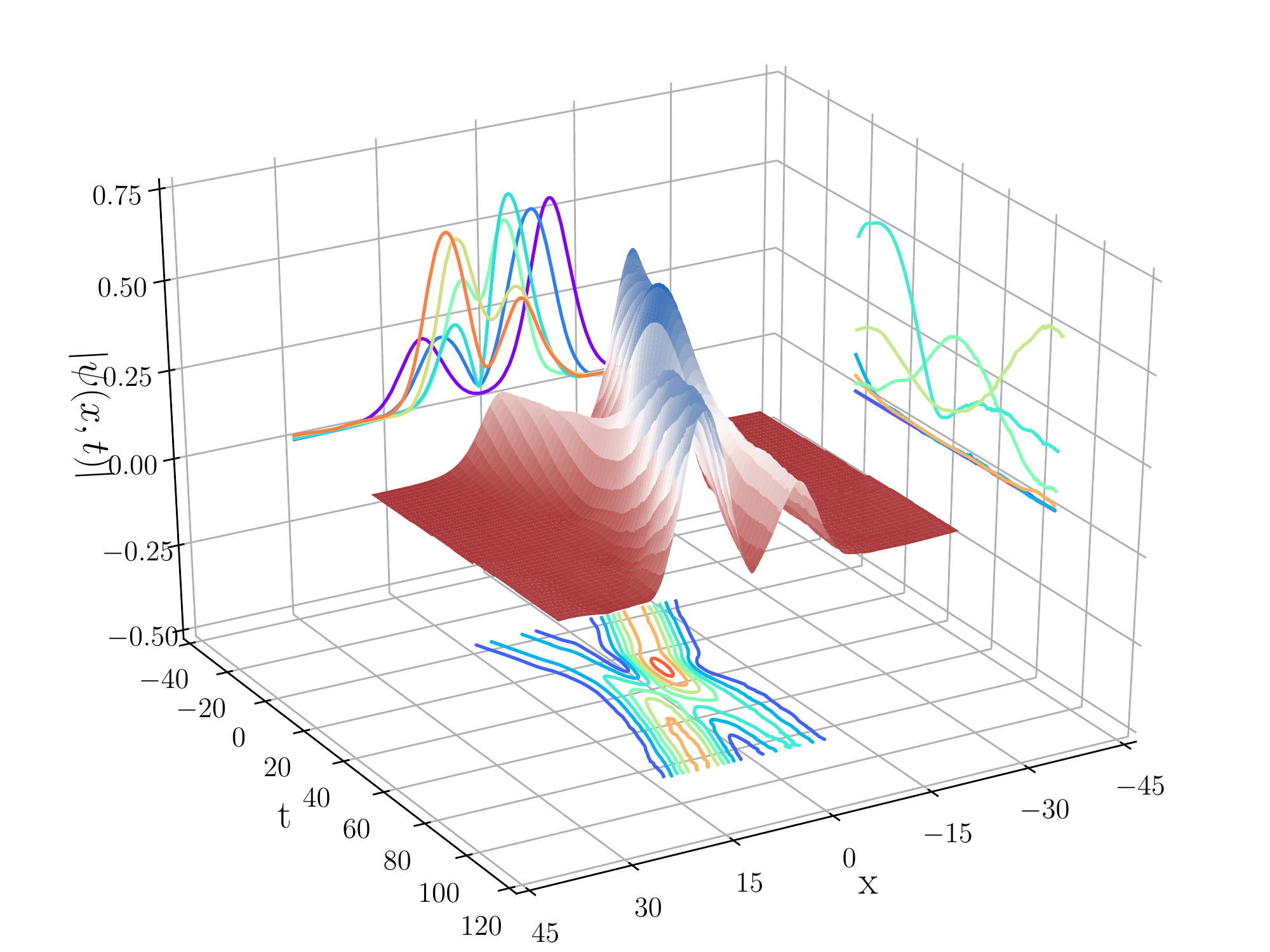}
\end{minipage}%
}%
\centering
\caption{(Color online) The training results for colliding droplets $\psi$ of third dynamic behaviors arising from the tpPINNs with 15200 initial and boundary points (cyan $``\times"$ in learned dynamics): (a) The reference, learned dynamics and error dynamics density plots, as well as sectional drawings which contain the learned and reference colliding droplet $\psi$ at five distinct times $t=15, 30, 45, 60$ and 75 (black dashed lines in density plots); (b) The 3D plot with contour map for the data-driven colliding droplets.}
\label{F7}
\end{figure}

Likewise, we still utilize 15200 initial and boundary points in tpPINNs, then the relative $L^2$ norm error reaches 6.449731$\rm e$-02. The training results of the data-driven colliding droplets stemming from the tpPINNs are indicated in Fig. \ref{F7}, in which Fig. \ref{F7}(a) displays the abundant density plots and sectional drawings, while Fig. \ref{F7}(b) showcases the 3D plot and its contour map on three planes. Compared with the second dynamic behavior of colliding droplets, the colliding droplets in this section produce obvious phase shift after droplets collision owing to the effect of relative phase $\varphi$. Furthermore, due to the introduction of relative phase, at a certain same moment, the amplitude of one droplet in Fig. \ref{F7}(a) is higher than that in Fig. \ref{F6}(a). The phase difference of some value is known to effectively induce repulsion between solitons \cite{Nguyen-NP2014}, thus the amplitude attenuation of two droplets after collision is not obvious, which is different the case of second dynamic behaviors. The evolution curve figures of the loss function arising from the tpPINNs with 9 stage subnets are presented in Fig \ref{F-loss5} [see Appendix \ref{App-B}].

\subsection{Breather excitations on droplet background}

Owing to the influence of quantum fluctuation, the introduction of LHY term made the self bound of condensates possible so admitted the presence of QDs. The integrability of the GPE with LHY correction term is destroyed, which hinders our study of the exact solution of the breather. Thus, different from previous subsections, we consider how to excite the numerical breather of the model through a small number of initial and boundary data points in the tpPINNs. Recently, the authors obtained the condition of breather excitations by means of the linear stability analysis in the GPE model with the LHY correction term, and excited the Akhmediev breathers and Kuznetsov-Ma breathers on the droplet by adding the periodic and localized modulations to nonlinear strength after quenching, respectively \cite{Lv-PLA2022}. Similarly, in this part, we focus on breather excitation on perturbed homogeneous BECs from the initial wave function with the homogeneous background wave and perturbations. The initial wave function [initial condition] is written as follows
\begin{align}\label{BE-iwf3-ic}
\psi(x,t=0)=\psi_0[1+AF(x)\mathrm{e}^{\mathrm{i}\theta}],\,x\in[x_{\mathrm{lb}},x_{\mathrm{ub}}],
\end{align}
in which $\psi_0(x,0)=a\rm{e}^{\mathrm{i}\nu x}$ is initial homogeneous background, while $F(x)$ is arbitrary initial perturbation with perturbation wave number. $A$ is the perturbation amplitude, and $\theta$ is the perturbation phase. Moreover, we utilize the following periodic boundary conditions,
\begin{align}\nonumber
\psi(x_{\mathrm{lb}},t)=\psi(x_{\mathrm{ub}},t),\,\psi_x(x_{\mathrm{lb}},t)=\psi_x(x_{\mathrm{ub}},t),\,t\in[0,T].
\end{align}
thus we obtain the formulation of periodic boundary conditions data loss
\begin{align}\label{BE-rbc}
\mathcal{L}_{\rm{bd}}(\mathcal{D}_{\rm{b}};\bm{\theta})=\frac{1}{N^k_{\rm{b}}}\Big[\big\|\bm{q}^{\bm{\theta},\mathrm{lb}}-\bm{q}^{\bm{\theta},\mathrm{ub}}\big\|^2_2+\big\|\bm{q}_x^{\bm{\theta},\mathrm{lb}}-\bm{q}_x^{\bm{\theta},\mathrm{ub}}\big\|^2_2\Big].
\end{align}
Next, by means of tpPINNs, we consider using periodic perturbation of cosine type [$F(x)=\cos(\delta x)$] to excite the breather on droplet background.

We set the parameters as $G=3$ [see GPE \eqref{MS-GPE-dm}], $A=0.1, \delta=0.8$ and $\theta=0.4\pi$. Then we provide the hyper-parameters and training information in tpPINNs as shown in table \ref{Tab3.3-1} [see Appendix \ref{App-A}]. After that, by using the initial-boundary conditions \eqref{BE-iwf3-ic} and \eqref{SD-iwf1-bc} in the spatiotemporal region $[0,4]\times[-20,20]$, we succeeded in exciting interesting data-driven Akhmediev breathers on droplet background wave.

\begin{figure}[htbp]
\centering
\subfigure[]{
\begin{minipage}[t]{0.48\textwidth}
\centering
\includegraphics[height=4.5cm,width=7cm]{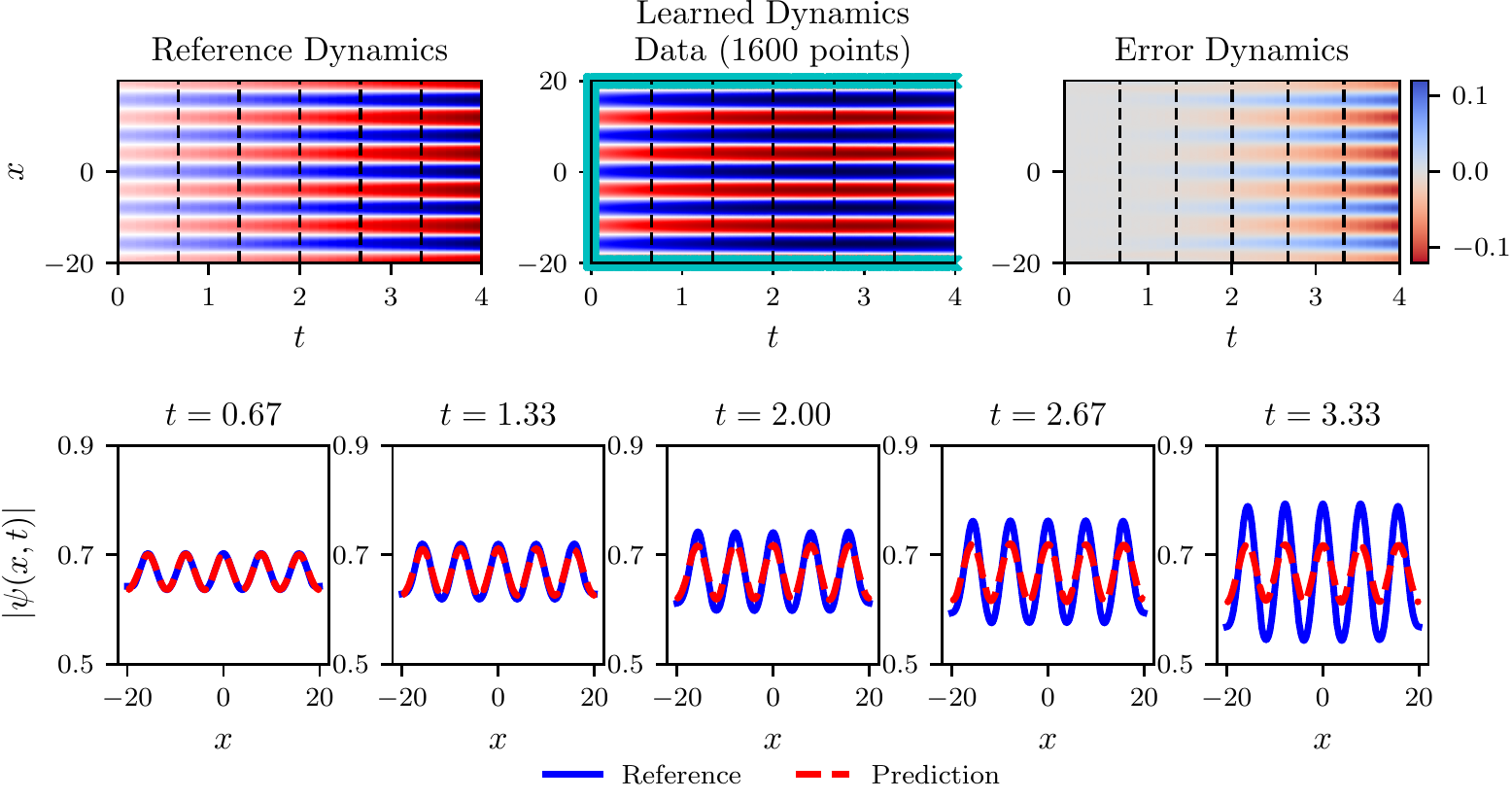}
\end{minipage}
}%
\subfigure[]{
\begin{minipage}[t]{0.48\textwidth}
\centering
\includegraphics[height=5cm,width=7cm]{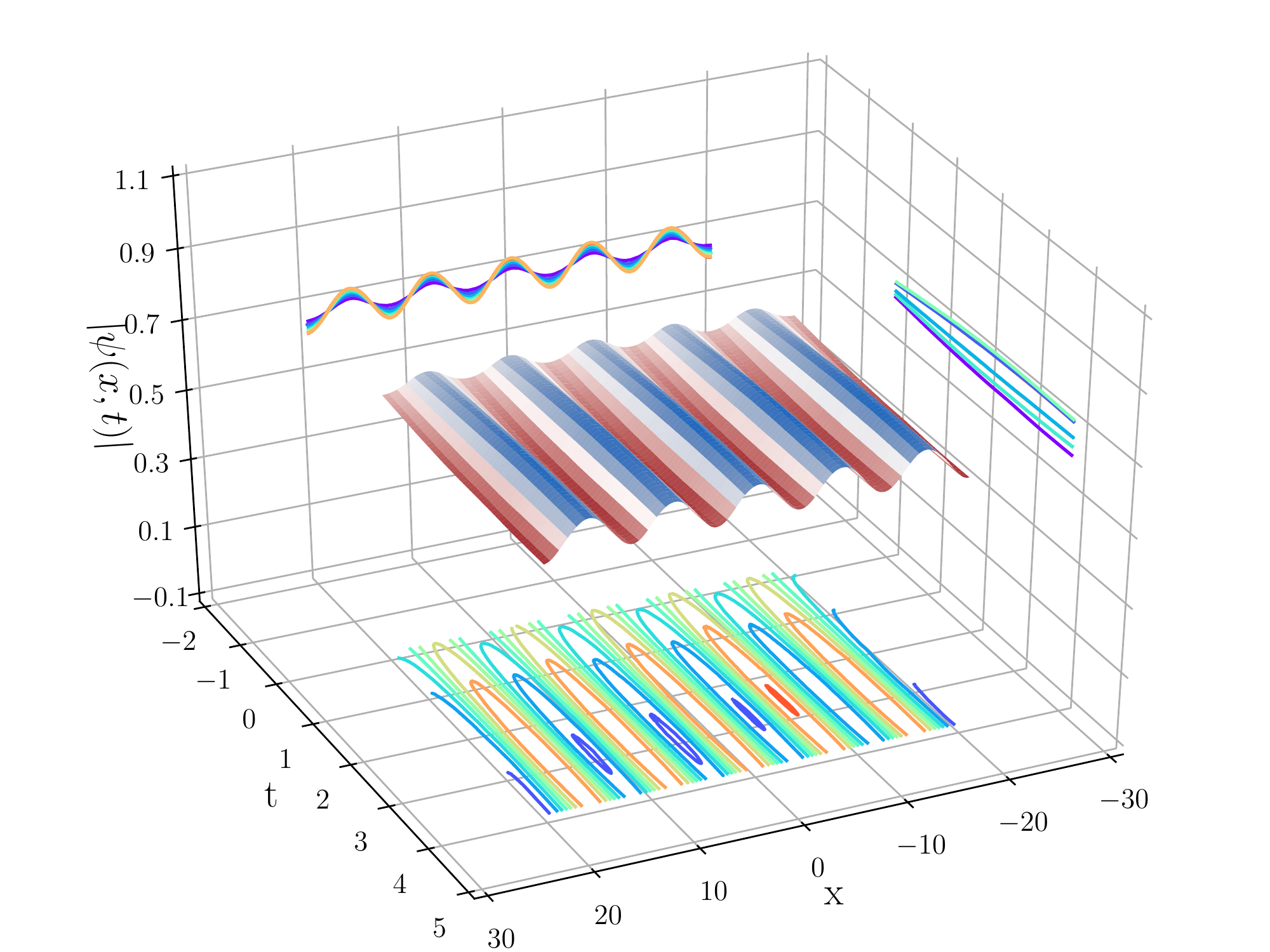}
\end{minipage}%
}%
\centering
\caption{(Color online) The training results for Akhmediev breathers excitations on droplet background arising from the tpPINNs with 1600 initial and boundary points (cyan $``\times"$ in learned dynamics): (a) The reference, learned dynamics and error dynamics density plots, as well as sectional drawings which contain the learned and reference Akhmediev breathers $\psi$ at five distinct moments $t=0.67, 1.33, 2.00, 2.67$ and 3.33 (black dashed lines in density plots); (b) The 3D plot with contour map for the data-driven Akhmediev breathers.}
\label{F8}
\end{figure}

After selecting 1600 initial and boundary points, then the relative $L^2$ norm error results from tpPINNs reaches 4.911635$\rm{e}$-02. Fig. \ref{F8} displays the training results of data-driven Akhmediev breathers arising from the tpPINNs. Fig. \ref{F8}(a)  manifests the abundant density plots and sectional drawings, and Fig. \ref{F8}(b) exhibits the 3D plot and their contour maps on three planes. As one can see from the plots, with time progresses (after time $t>2$), the $L^2$ norm error of data-driven Akhmediev breathers becomes larger and larger, which is mainly concentrated in the amplitude. Interestingly, different from the traditional Akhmediev breather (with only one row of peaks), the data-driven Akhmediev breathers with double-row peaks is formed on the plane wave owing to the interplay between periodic perturbation and homogeneous background droplet, as shown in \ref{F8}. From Ref. \cite{Lv-PLA2022}, when increasing $\theta$, Akhmediev breathers with double-row peaks become Akhmediev breathers with single-row peaks, whose frequency also increases accordingly. Furthermore, the evolution curve plots of the loss function in tpPINNs are displays in Fig \ref{F-loss6} [see Appendix \ref{App-B}].

Exciting breathers on a plane wave background in integrable systems has long been a research focus. However, in non-integrable systems, breather excitation on a background wave can be more challenging owing to its dynamic behavior may become very complex when the parameters are slightly changed. In this subsection, although we can set more diverse functions $F(x)$,  certain functions such as $F(x)=\mathrm{sech}(0.2x)$ may generate high-frequency dynamic behaviors that can present challenges for deep learning.

\section{Conclusions}

In this work, we propose a novel PINN model to study complex dynamics on the one-dimensional QDs in BECs. Owing to the dynamic behaviors of QDs needs to be studied in a long time domain, and the simulation effect of the traditional PINN on the numerical solution in a long time domain is not satisfactory, thus we propose a tpPINNs, which is feasible to use it to study the complex dynamics of QDs. For segmenting the long time domain, we introduce the concept of pseudo initial point, and showcase the example of taking points in each time piecewise domain [total $n$ stages], which is displayed in detail in Fig. \ref{F1}. The schematic architecture of tpPINNs mosel is displayed in Fig. \ref{F2}, mainly including three parts. Moreover, dimensionless GPE model can be derived by using specific transformation, then the ground-state solution and its special asymptotic behavior of GPE are revealed. By using tpPINNs and providing different initial and boundary data of GPE, we showcase the abundant dynamic behaviors of one-dimensional QDs. In the case of intrinsic modulation of single droplet, we train two completely opposite dynamic behaviors of QDs in tpPINNs by adjusting wave number $\kappa$, namely droplet avoiding fragmentation and droplet splitting. In the case of collision between two droplets, we learn three interesting dynamic behaviors by choosing different parameters in initial condition. During the collision process of the first dynamic behaviors, we learn an interference pattern resulting from tpPINNs, which is a significant feature of the interplay of coherent matter waves. Moreover, we excite the breather on droplet background by using initial data with periodic perturbation of cosine type. Our study demonstrates the ability of tpPINNs to capture the rich and complex dynamics of one-dimensional QDs in BECs. By providing different initial and boundary data for GPE, we showcase intrinsic modulation of droplets, droplet collisions interplay, and excitation of breathers on droplet background. This provides important guidance for future research on QDs in BECs utilizing deep learning technology.

\section*{Declaration of competing interest}
The authors declare that they have no known competing financial interests or personal relationships that could have appeared to influence the work reported in this paper.

\section*{Acknowledgements}
\hspace{0.3cm}
The authors gratefully acknowledge the support of the National Natural Science Foundation of China (No. 12175069 and No. 12235007), Science and Technology Commission of Shanghai Municipality (No. 21JC1402500 and No. 22DZ2229014), and Natural Science Foundation of Shanghai (No. 23ZR1418100).

\section*{Appendix}

\appendix
\renewcommand{\appendixname}{Appendix~\Alph{section}}

\section{The training informations of tpPINNs}\label{App-A}

The training informations of tpPINNs with different number of subnet are shown as following tables \ref{Tab1}-\ref{Tab3.3-1}.
\begin{table}[htbp]
  \caption{Training information of tpPINNs ($k=15$ subnets) in case of droplet avoids fragmentation}
  \label{Tab1}
  \centering
  \scalebox{0.8}{
  \begin{tabular}{l|c|c|c|c|c|c|c|c}
  \toprule
  \diagbox{\small{\textbf{Networks}}}{\small{\textbf{Types}}} & $L,\,N_d$ & $t$ & $N^k_{\rm{i}}$ & $N^k_{\rm{b}}$ & $N^k_{\rm{c}}$ & $\alpha,\,\beta$ & Adam & $L^2$ error \\
  \hline
  1-st stage   & 5,\,200 & [0,10]  & 1000 & 150 & 50000 & $0.00001,\,0.1$ & 50000 & 1.310815$\rm e$-02 \\
  \hline
  2-nd stage   & 5,\,200 & [10,20] & 1000 & 150 & 60000 & $0.00001,\,0.1$ & 50000 & 2.786412$\rm e$-02 \\
  \hline
  3-rd stage   & 5,\,200 & [20,30] & 1000 & 150 & 60000 & $0.00001,\,0.1$ & 50000 & 4.154131$\rm e$-02 \\
  \hline
  4-th stage   & 5,\,200 & [30,40] & 1000 & 150 & 60000 & $0.00001,\,0.1$ & 50000 & 4.913788$\rm e$-02 \\
  \hline
  5-th stage   & 5,\,200 & [40,50] & 1000 & 150 & 60000 & $0.00001,\,0.1$ & 50000 & 5.017800$\rm e$-02 \\
  \hline
  6-th stage   & 5,\,200 & [50,60] & 1000 & 150 & 60000 & $0.00001,\,0.1$ & 50000 & 4.826475$\rm e$-02 \\
  \hline
  7-th stage   & 5,\,200 & [60,70] & 1000 & 150 & 80000 & $0.00001,\,0.1$ & 50000 & 4.912388$\rm e$-02 \\
  \hline
  8-th stage   & 5,\,200 & [70,80] & 1000 & 150 & 80000 & $0.00001,\,0.1$ & 50000 & 4.889775$\rm e$-02 \\
  \hline
  9-th stage   & 5,\,200 & [80,90] & 1000 & 150 & 80000 & $0.00001,\,0.1$ & 50000 & 5.323550$\rm e$-02 \\
  \hline
  10-th stage  & 5,\,200 & [90,100] & 1000 & 150 & 80000 & $0.00001,\,0.1$ & 50000 & 6.425244$\rm e$-02 \\
  \hline
  11-th stage  & 5,\,200 & [100,110] & 1000 & 150 & 80000 & $0.00001,\,0.1$ & 50000 & 5.444058$\rm e$-02 \\
  \hline
  12-th stage  & 5,\,200 & [110,120] & 1000 & 150 & 80000 & $0.00001,\,0.1$ & 50000 & 6.656296$\rm e$-02 \\
  \hline
  13-th stage  & 5,\,200 & [120,130] & 1000 & 150 & 80000 & $0.00001,\,0.1$ & 50000 & 6.493988$\rm e$-02 \\
  \hline
  14-th stage  & 5,\,200 & [130,140] & 1000 & 150 & 80000 & $0.00001,\,0.1$ & 50000 & 6.812361$\rm e$-02 \\
  \hline
  15-th stage  & 5,\,200 & [140,150] & 1000 & 150 & 80000 & $0.00001,\,0.1$ & 50000 & 7.057909$\rm e$-02\\
  \hline
  tpPINNs & \diagbox{}{} & [0,150] & 1000 & 2250 & \diagbox{}{} & \diagbox{}{} & \diagbox{}{} & 5.362393$\rm e$-02 \\
  \bottomrule
  \end{tabular}}
\end{table}

\begin{table}[htbp]
  \caption{Training information of tpPINNs ($k=5$ subnets) in case of droplet splitting}
  \label{Tab2}
  \centering
  \scalebox{0.8}{
  \begin{tabular}{l|c|c|c|c|c|c|c|c}
  \toprule
  \diagbox{\small{\textbf{Networks}}}{\small{\textbf{Types}}} & $L,\,N_d$ & $t$ & $N^k_{\rm{i}}$ & $N^k_{\rm{b}}$ & $N^k_{\rm{c}}$ & $\alpha,\,\beta$ & Adam & $L^2$ error \\
  \hline
  1-st stage   & 5,\,200 & [0,5]  & 1000 & 50 & 50000 & $0.00001,\,1$ & 50000 & 1.434686$\rm e$-02 \\
  \hline
  2-nd stage   & 5,\,200 & [5,10] & 1000 & 50 & 60000 & $0.00001,\,1$ & 50000 & 2.575081$\rm e$-02 \\
  \hline
  3-rd stage   & 5,\,200 & [10,15] & 1000 & 50 & 60000 & $0.00001,\,1$ & 50000 & 4.166505$\rm e$-02 \\
  \hline
  4-th stage   & 5,\,200 & [15,20] & 1000 & 50 & 60000 & $0.00001,\,1$ & 50000 & 6.581496$\rm e$-02 \\
  \hline
  5-th stage   & 5,\,200 & [20,25] & 1000 & 50 & 60000 & $0.00001,\,1$ & 50000 & 1.316989$\rm e$-01 \\
  \hline
  tpPINNs & \diagbox{}{} & [0,25] & 1000 & 250 & \diagbox{}{} & \diagbox{}{} & \diagbox{}{} & 6.993923$\rm e$-02 \\
  \bottomrule
  \end{tabular}}
\end{table}

\begin{table}[htbp]
  \caption{Training information of tpPINNs ($k=4$ subnets) in first case of colliding droplets}
  \label{Tab3.2-1}
  \centering
  \scalebox{0.8}{
  \begin{tabular}{l|c|c|c|c|c|c|c|c}
  \toprule
  \diagbox{\small{\textbf{Networks}}}{\small{\textbf{Types}}} & $L,\,N_d$ & $t$ & $N^k_{\rm{i}}$ & $N^k_{\rm{b}}$ & $N^k_{\rm{c}}$ & $\alpha,\,\beta$ & Adam & $L^2$ error \\
  \hline
  1-st stage   & 5,\,200 & [0,2.5]  & 800 & 400 & 50000 & $0.0001,\,1$ & 50000 & 2.567498$\rm e$-02 \\
  \hline
  2-nd stage   & 5,\,200 & [2.5,5.0] & 800 & 400 & 70000 & $0.0001,\,1$ & 50000 & 3.963029$\rm e$-02 \\
  \hline
  3-rd stage   & 5,\,200 & [5.0,7.5] & 800 & 400 & 70000 & $0.0001,\,1$ & 50000 & 6.310087$\rm e$-02 \\
  \hline
  4-th stage   & 5,\,200 & [7.5,10] & 800 & 400 & 70000 & $0.0001,\,1$ & 50000 & 8.823686$\rm e$-02 \\
  \hline
  tpPINNs & \diagbox{}{} & [0,10] & 800 & 1600 & \diagbox{}{} & \diagbox{}{} & \diagbox{}{} & 5.919966$\rm e$-02 \\
  \bottomrule
  \end{tabular}}
\end{table}

\begin{table}[htbp]
  \caption{Training information of tpPINNs ($k=9$ subnets) in second case of colliding droplets}
  \label{Tab3.2-3}
  \centering
  \scalebox{0.8}{
  \begin{tabular}{l|c|c|c|c|c|c|c|c}
  \toprule
  \diagbox{\small{\textbf{Networks}}}{\small{\textbf{Types}}} & $L,\,N_d$ & $t$ & $N^k_{\rm{i}}$ & $N^k_{\rm{b}}$ & $N^k_{\rm{c}}$ & $\alpha,\,\beta$ & Adam & $L^2$ error \\
  \hline
  1-st stage   & 5,\,200 & [0,10]  & 800 & 800 & 50000 & $0.0001,\,1$ & 50000 & 9.588465$\rm e$-03 \\
  \hline
  2-nd stage   & 5,\,200 & [10,20] & 800 & 800 & 70000 & $0.0001,\,1$ & 50000 & 2.279675$\rm e$-02 \\
  \hline
  3-rd stage   & 5,\,200 & [20,30] & 800 & 800 & 70000 & $0.0001,\,1$ & 50000 & 4.216680$\rm e$-02 \\
  \hline
  4-th stage   & 5,\,200 & [30,40] & 800 & 800 & 70000 & $0.0001,\,1$ & 50000 & 5.955930$\rm e$-02 \\
  \hline
  5-th stage   & 5,\,200 & [40,50] & 800 & 800 & 70000 & $0.0001,\,1$ & 50000 & 6.550802$\rm e$-02 \\
  \hline
  6-th stage   & 5,\,200 & [50,60] & 800 & 800 & 80000 & $0.00001,\,1$ & 50000 & 7.335074$\rm e$-02 \\
  \hline
  7-th stage   & 5,\,200 & [60,70] & 800 & 800 & 80000 & $0.00001,\,1$ & 50000 & 9.111827$\rm e$-02 \\
  \hline
  8-th stage   & 5,\,200 & [70,80] & 800 & 800 & 80000 & $0.00001,\,1$ & 50000 & 1.143625$\rm e$-01 \\
  \hline
  9-th stage   & 5,\,200 & [80,90] & 800 & 800 & 80000 & $0.00001,\,1$ & 50000 & 1.469812$\rm e$-01 \\
  \hline
  tpPINNs & \diagbox{}{} & [0,90] & 800 & 7200 & \diagbox{}{} & \diagbox{}{} & \diagbox{}{} & 8.071953$\rm e$-02 \\
  \bottomrule
  \end{tabular}}
\end{table}

\begin{table}[htbp]
  \caption{Training information of tpPINNs ($k=9$ subnets) in third case of colliding droplets}
  \label{Tab3.2-4}
  \centering
  \scalebox{0.8}{
  \begin{tabular}{l|c|c|c|c|c|c|c|c}
  \toprule
  \diagbox{\small{\textbf{Networks}}}{\small{\textbf{Types}}} & $L,\,N_d$ & $t$ & $N^k_{\rm{i}}$ & $N^k_{\rm{b}}$ & $N^k_{\rm{c}}$ & $\alpha,\,\beta$ & Adam & $L^2$ error \\
  \hline
  1-st stage   & 5,\,200 & [0,10]  & 800 & 800 & 50000 & $0.0001,\,1$ & 50000 & 6.140534$\rm e$-03 \\
  \hline
  2-nd stage   & 5,\,200 & [10,20] & 800 & 800 & 70000 & $0.0001,\,1$ & 50000 & 1.131305$\rm e$-02 \\
  \hline
  3-rd stage   & 5,\,200 & [20,30] & 800 & 800 & 70000 & $0.0001,\,1$ & 50000 & 1.700895$\rm e$-02 \\
  \hline
  4-th stage   & 5,\,200 & [30,40] & 800 & 800 & 70000 & $0.0001,\,1$ & 50000 & 2.251825$\rm e$-02 \\
  \hline
  5-th stage   & 5,\,200 & [40,50] & 800 & 800 & 70000 & $0.0001,\,1$ & 50000 & 2.774027$\rm e$-02 \\
  \hline
  6-th stage   & 5,\,200 & [50,60] & 800 & 800 & 80000 & $0.00001,\,1$ & 50000 & 5.317747$\rm e$-02 \\
  \hline
  7-th stage   & 5,\,200 & [60,70] & 800 & 800 & 80000 & $0.00001,\,1$ & 50000 & 8.095352$\rm e$-02 \\
  \hline
  8-th stage   & 5,\,200 & [70,80] & 800 & 800 & 80000 & $0.00001,\,1$ & 50000 & 1.019022$\rm e$-01 \\
  \hline
  9-th stage   & 5,\,200 & [80,90] & 800 & 800 & 80000 & $0.00001,\,1$ & 50000 & 1.261384$\rm e$-01 \\
  \hline
  tpPINNs & \diagbox{}{} & [0,90] & 800 & 7200 & \diagbox{}{} & \diagbox{}{} & \diagbox{}{} & 6.449731$\rm e$-02 \\
  \bottomrule
  \end{tabular}}
\end{table}

\begin{table}[htbp]
  \caption{Training information of tpPINNs ($k=3$ subnets) in case of breather excitations on droplet background}
  \label{Tab3.3-1}
  \centering
  \scalebox{0.8}{
  \begin{tabular}{l|c|c|c|c|c|c|c|c}
  \toprule
  \diagbox{\small{\textbf{Networks}}}{\small{\textbf{Types}}} & $L,\,N_d$ & $t$ & $N^k_{\rm{i}}$ & $N^k_{\rm{b}}$ & $N^k_{\rm{c}}$ & $\alpha,\,\beta$ & Adam & $L^2$ error \\
  \hline
  1-st stage   & 5,\,200 & [0,2]  & 600 & 300 & 30000 & $0.00005,\,10$ & 60000 & 9.229369$\rm e$-03 \\
  \hline
  2-nd stage   & 5,\,200 & [2,3] & 500 & 100 & 60000 & $0.00001,\,10$ & 50000 & 3.948447$\rm e$-02 \\
  \hline
  3-rd stage   & 5,\,200 & [3,4] & 400 & 100 & 70000 & $0.00003,\,10$ & 70000 & 8.879516$\rm e$-02 \\
  \hline
  tpPINNs & \diagbox{}{} & [0,4] & 600 & 500 & \diagbox{}{} & \diagbox{}{} & \diagbox{}{} & 4.911635$\rm e$-02 \\
  \bottomrule
  \end{tabular}}
\end{table}

\section{Evolution of the loss function in tpPINNs}\label{App-B}

We showcase the evolution curve of the loss function in tpPINNs by utilizing Adam optimizer in the following figures \ref{F-loss1}-\ref{F-loss5} [here Adam optimizer records relative $L^2$ norm error every 10 iterations].

\begin{figure}[htbp]
\centering
\includegraphics[height=7cm,width=15cm]{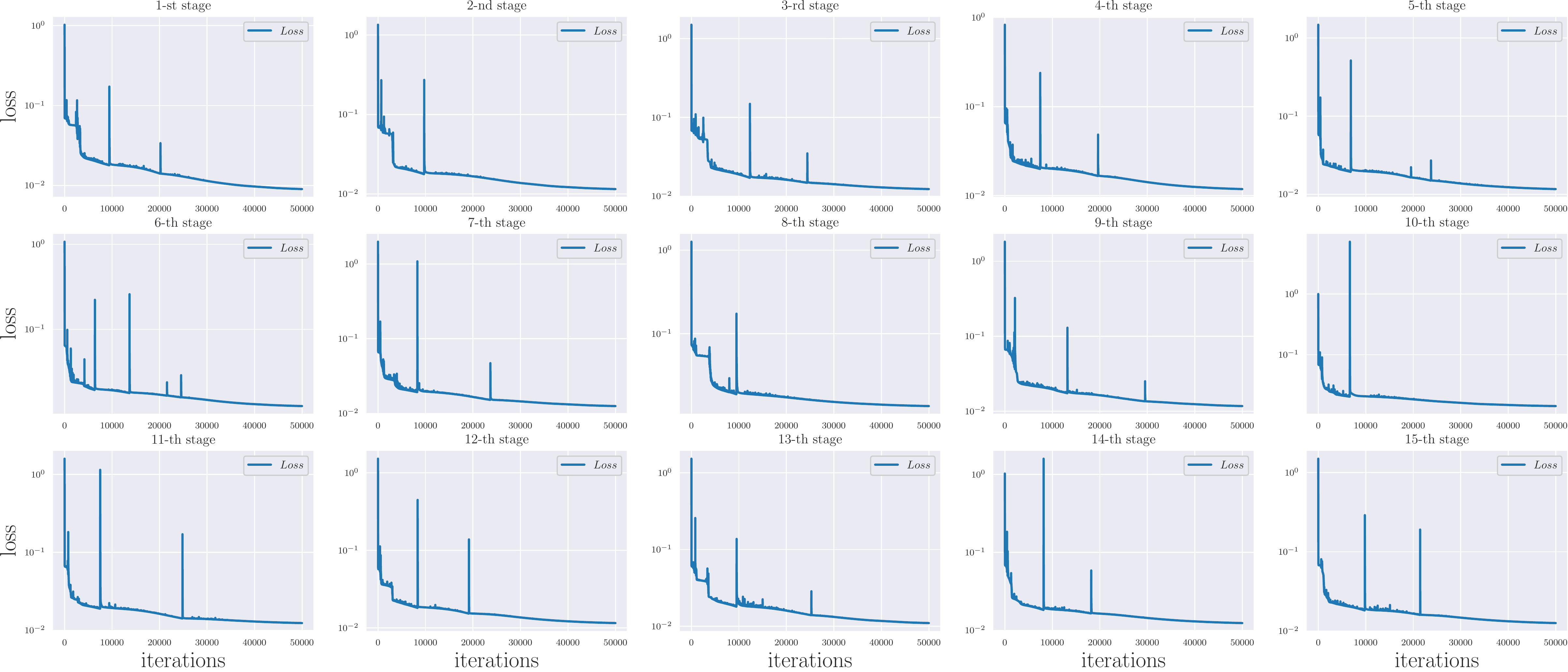}
\caption{(Color online) Evolution of the loss function in tpPINNs, as well as the relative $L^2$ norm error during training in case of droplet avoiding fragmentation.}
\label{F-loss1}
\end{figure}

\begin{figure}[htbp]
\centering
\includegraphics[height=2cm,width=15cm]{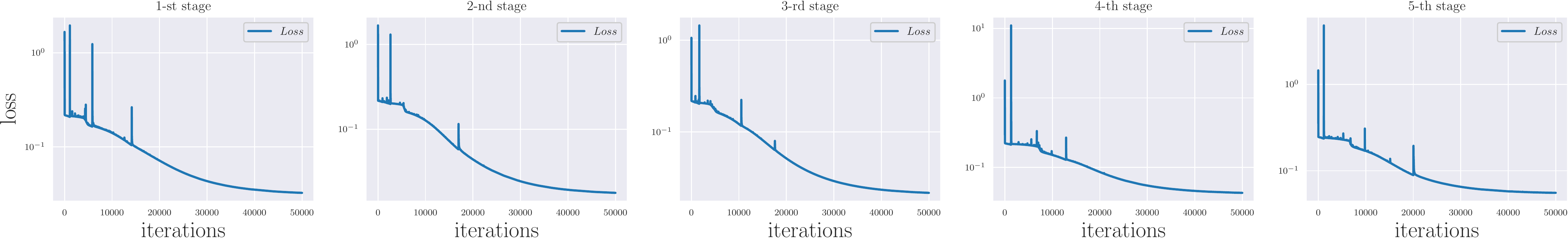}
\caption{(Color online) Evolution of the loss function in tpPINNs, as well as the relative $L^2$ norm error during training in case of droplet splitting.}
\label{F-loss2}
\end{figure}

\begin{figure}[htbp]
\centering
\includegraphics[height=3cm,width=15cm]{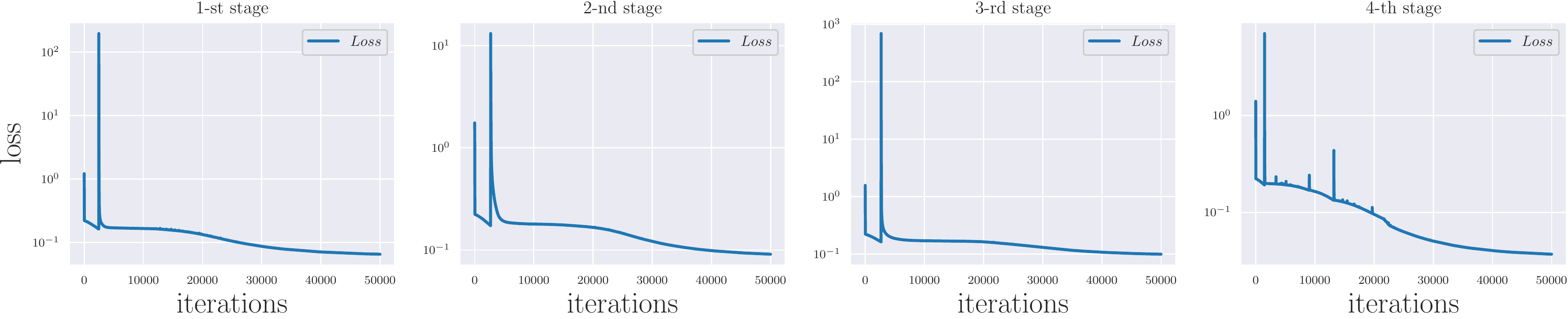}
\caption{(Color online) Evolution of the loss function in tpPINNs, as well as the relative $L^2$ norm error during training in first case of colliding droplets.}
\label{F-loss3}
\end{figure}

\begin{figure}[htbp]
\centering
\includegraphics[height=9.5cm,width=15cm]{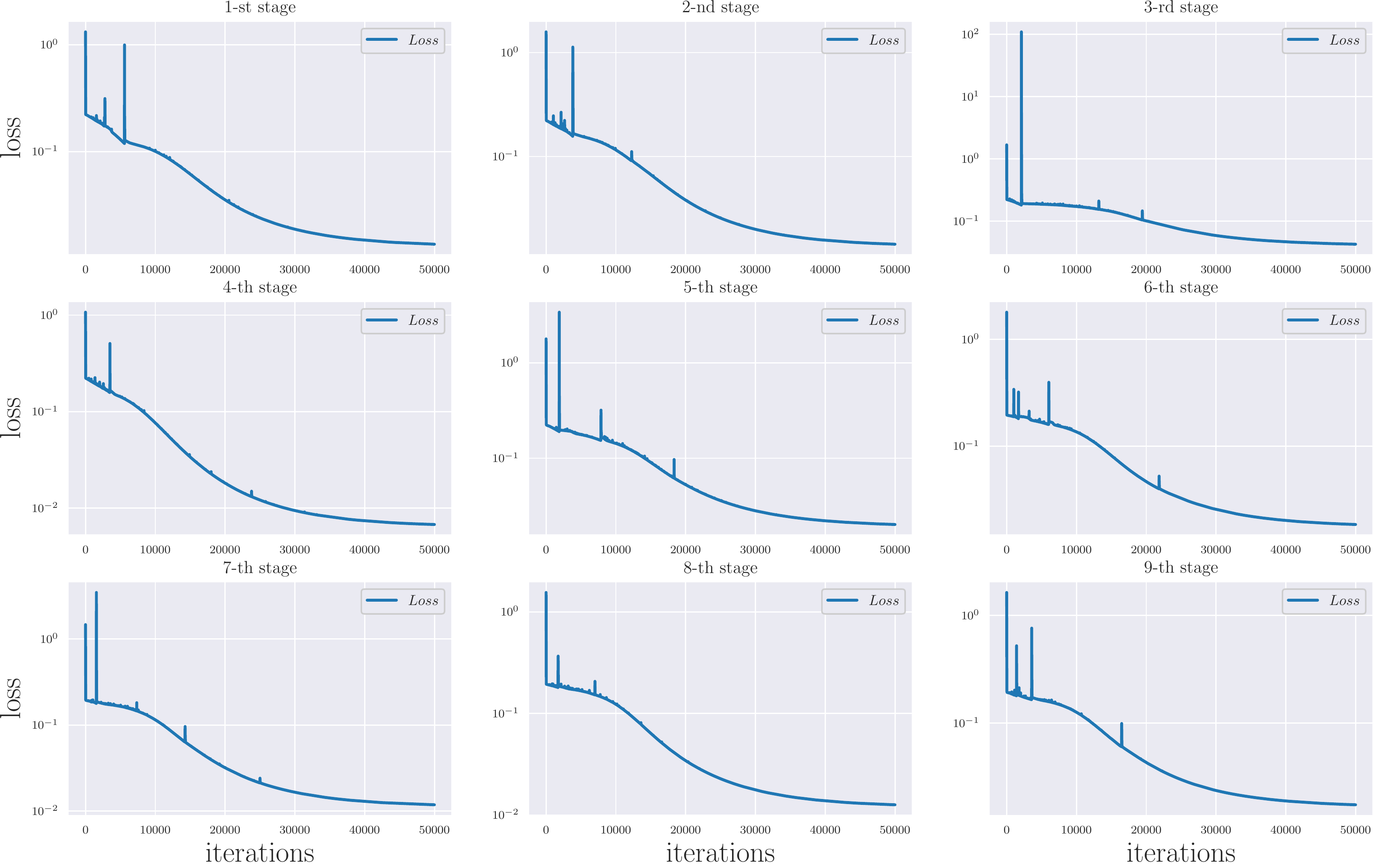}
\caption{(Color online) Evolution of the loss function in tpPINNs, as well as the relative $L^2$ norm error during training in second case of colliding droplets.}
\label{F-loss4}
\end{figure}

\begin{figure}[htbp]
\centering
\includegraphics[height=9.5cm,width=15cm]{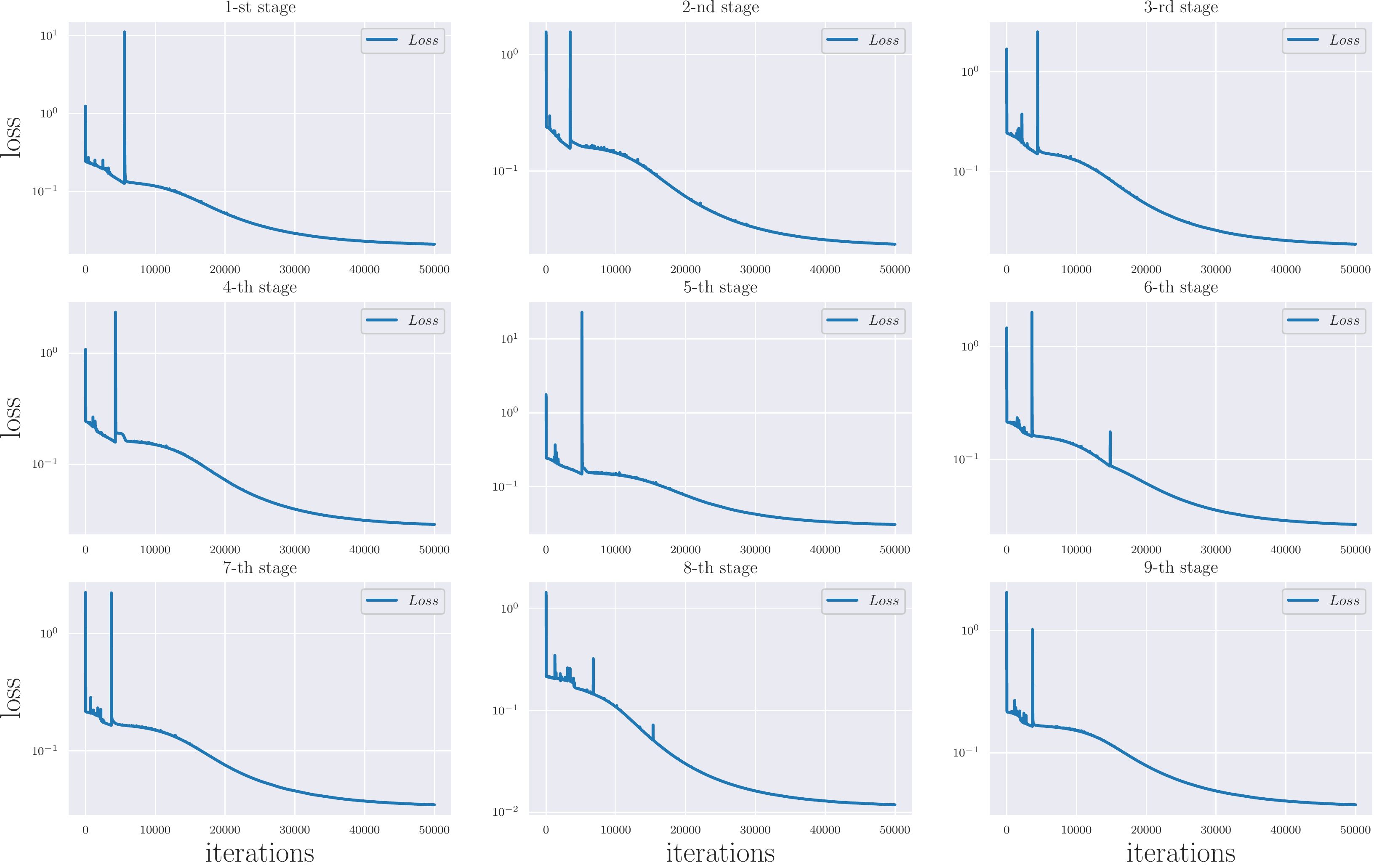}
\caption{(Color online) Evolution of the loss function in tpPINNs, as well as the relative $L^2$ norm error during training in third case of colliding droplets.}
\label{F-loss5}
\end{figure}

We showcase the evolution curve of the loss function in tpPINNs by utilizing Adam optimizer in the following figure \ref{F-loss6} [here Adam optimizer records relative $L^2$ norm error every 1 iterations].

\begin{figure}[htbp]
\centering
\includegraphics[height=4cm,width=13cm]{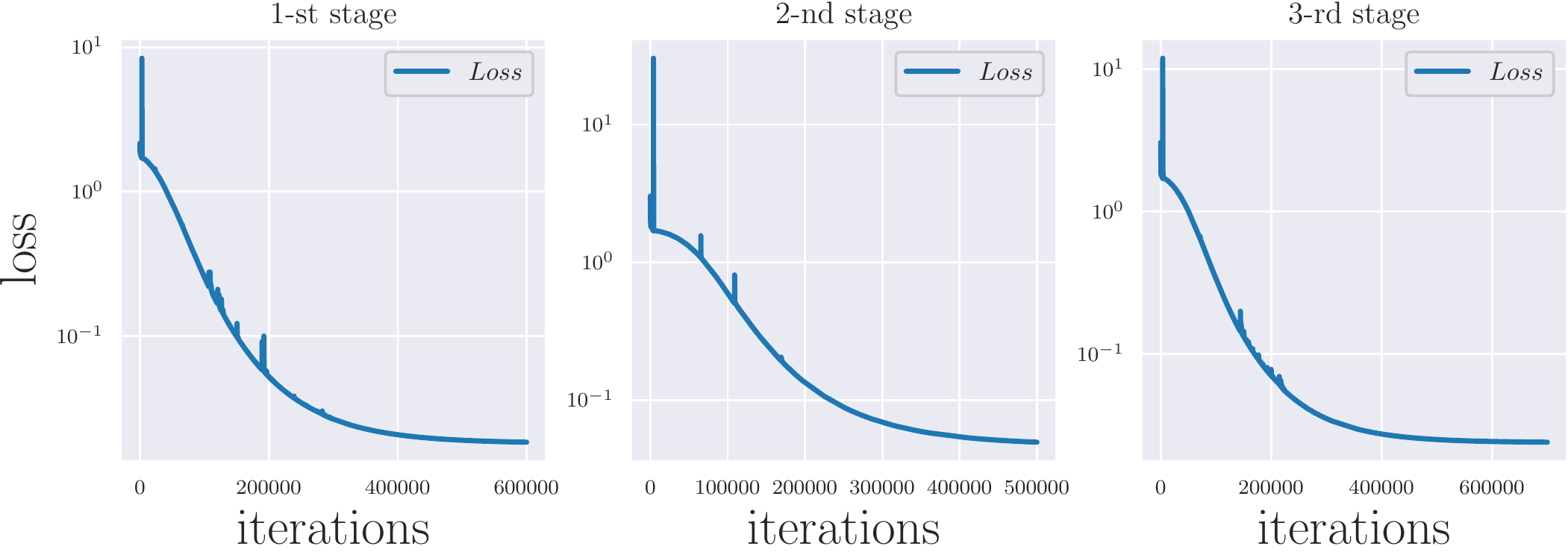}
\caption{(Color online) Evolution of the loss function in tpPINNs, as well as the relative $L^2$ norm error during training in case of breather excitations on droplet background.}
\label{F-loss6}
\end{figure}

\end{document}